\documentclass[preprint,12pt,authoryear]{elsarticle}

\usepackage{amssymb}
\usepackage{listings}
\usepackage{multirow}
\usepackage{microtype}
\usepackage{comment}
\usepackage{hyperref}
\usepackage{cleveref}

\def\baseCommit END{\emph{Base}}
\def\leftCommit END{\emph{Left}}
\def\rightCommit END{\emph{Right}}
\def\mergeCommit END{\emph{Merge}}

\newcommand{\rev}[1]{\textcolor{black}{#1}}

\usepackage{xcolor}
\definecolor{excerptgreen}{HTML}{AAEEB7}
\definecolor{excerptred}{HTML}{E4A598}
\begin{document}
\definecolor{light-gray-fork}{gray}{0.925}
\definecolor{light-gray-mainline}{gray}{0.8}
\definecolor{javared}{rgb}{0.6,0,0} % for strings
\definecolor{javagreen}{rgb}{0.25,0.5,0.35} % comments
\definecolor{javapurple}{rgb}{0.5,0,0.35} % keywords
\definecolor{javadocblue}{rgb}{0.25,0.35,0.75} % javadoc

\begin{frontmatter}



\title{Detecting Semantic Conflicts with Unit Tests}

\author[label1]{Léuson Da Silva}
\author[label1]{Paulo Borba}
\author[label1]{Toni Maciel}
\author[label2,label3]{Wardah Mahmood}
\author[label2,label4]{Thorsten Berger}
\author[label1]{João Moisakis}
\author[label1]{Aldiberg Gomes}
\author[label1]{Vinícius Leite}

\affiliation[label1]{organization={Informatics Center, Federal University of Pernambuco},
             state={Pernambuco},
             country={Brazil}}

 \affiliation[label2]{organization={Chalmers University},
             city={Gothenburg},
             country={Sweden}}

\affiliation[label3]{organization={University of Gothenburg},
             city={Gothenburg},
             country={Sweden}}

\affiliation[label4]{organization={Ruhr University Bochum},
             city={Bochum},
             country={Germany}}

\begin{abstract}
Branching and merging are common practices in collaborative software development. They increase developer productivity, allowing developers to independently contribute to a software project. Despite such benefits, these practices come at a cost--- the need to merge software and resolve merge conflicts, which often occur in practice. While modern merge techniques, such as 3-way and structured merge, can resolve textual conflicts automatically, they fail when the conflict arises not at the syntactic but at the semantic level. Detecting such semantic conflicts requires understanding the behavior of the software, which is beyond the capabilities of most existing merge tools. Although semantic merge tools have been proposed, they are usually based on heavyweight static analyses, or need explicit specifications of program behavior. In this work, we take a different route and propose SAM (SemAntic Merge), a semantic merge tool based on the automated generation of unit tests that are used as partial specifications of the changes to be merged, and drive the detection of unwanted behavior changes (conflicts) when merging software. To evaluate SAM’s feasibility for detecting conflicts, we perform an empirical study relying on a dataset of more than 80 pairs of changes integrated to common class elements (constructors, methods, and fields) from 51 merge scenarios. We also assess how the four unit-test generation tools used by SAM individually contribute to conflict identification: EvoSuite (the standard and the differential version), Randoop, and Randoop Clean, an extended version of Randoop proposed here. Additionally, we propose and assess the adoption of Testability Transformations, which are changes directly applied to the code under analysis aiming to increase its testability during test suite generation, and Serialization, which aims to support unit-test tools to generate tests that manipulate complex objects. Our results show that SAM best performs when combining only the tests generated by Differential EvoSuite and EvoSuite, and using the proposed Testability Transformations (nine detected conflicts out of 28). These results reinforce previous findings about the potential of using test-case generation to detect test conflicts as a method that is versatile and requires only limited deployment effort in practice.
\end{abstract}



\begin{keyword}
Semantic Conflicts \sep Differential Testing \sep Behavior Change



\end{keyword}

\end{frontmatter}

\section{Introduction}
\label{sec:introduction}
\noindent
\looseness=-1
\noindent
Branching and merging are common practices in collaborative software development.
They facilitate effective teamwork, allowing developers to independently contribute to the same project.
Still, branching and merging come with costs, including the need to resolve conflicts that are detected by merge tools when integrating code changes.
Depending on project characteristics~\citep{predicting-merge-conflicts-KareshkNadiRubin2019,predictive-factors-merge-conflicts-DiasBorba2020}, such \emph{merge} conflicts often occur~\citep{parallel-changes-sw-development-PerrySiyVotta2001,merge-survey-Mens2002,mininig-cvs-commitsZimmermann2007,value-of-branchesBirdZimmermann2012,conflict-minimization-task-schedulingKasiSarma2013,early-detection-conflicts-speculativeBrunReidErnstNotkin2013,wardah2020elasticsearch}, even when using more advanced merge tools~\citep{semistructured-mergeApelLiebigBrandlLengauerKastner2011,structured-mergeApelLesenichLengauer2012,improved-semistructured-mergeCavalcantiBorbaAccioly2017,semistructure-merge-conflictsAcciolyBorbaCavalcanti2017,semistructured-vs-structured-mergeCavalcantiBorbaSeibtApel2019,semistructured-merge-jsTavaresBorbaCavalcantiSoares2019,refactoring-aware-structured-mergeShenZhangZhaoLiangJinWang2019} that explore language syntax and static semantics to avoid spurious conflicts. 

While many \emph{merge} conflicts are easy to fix, some of them can only be fixed with significant effort and knowledge of the code changes to be merged.
This can negatively affect development productivity, and even compromise software quality in case developers incorrectly fix conflicts~\citep{early-detection-conflicts-conservativeSarmaRedmilesHoek2012,value-of-branchesBirdZimmermann2012,merge-conflicts-perceptionMcKeeNelsonSarmaDig2017}. 
To avoid dealing with \emph{merge} conflicts, developers sometimes even adopt risky practices, such as rushing to finish changes first~\citep{collaborative-sw-developmentGrinter1996,early-detection-conflicts-conservativeSarmaRedmilesHoek2012} and partial check-ins~\citep{collaborative-development-private-publicdeSouzaRedmilesDourish2003}.
Similarly, partially motivated by the need to reduce \emph{merge} conflicts, development teams have been adopting techniques such as trunk-based development~\citep{release-engineeringAdamsMcIntosh2016,monorepo-googlePotvinLevenberg2016,sw-engineering-at-googleHenderson2017} and feature toggles~\citep{feature-toggles-devopsBassWeberZhu2015,release-engineeringAdamsMcIntosh2016,feature-toggleFowler2016,feature-flags-vs-branchingHodgson2017}.

Although these practices might reduce the occurrence of \emph{merge} conflicts, there is no evidence that they are effective in resolving or even detecting
%semantic conflicts. Such conflicts do not violate the syntax or static semantics of programming languages, but the dynamic semantics of the programs, which can only be detected at
 \emph{test}~\citep{early-detection-conflicts-speculativeBrunReidErnstNotkin2013} and \emph{production} conflicts, which are only observed when running project tests and using the system. 
 As such, they are more serious than \emph{merge} conflicts, because they give rise to software failures.
%\tb{I was struggling with that here, for two reasons: first, the prev. paragraph explains that resolving conflicts is laborious, so one thinks that the main problem addressed is performance/efficiency; second, I think one should talk about semantic conflicts, which are (partly) unwanted behavior changes (violating some, perhaps implicit, specification) and are observable at test time or production time}\pb{Thorsten, please check now, before and after this comment.}
In fact, some of the practices mentioned before might even aggravate the costs of test and production conflicts, which are special kinds of what we hereafter call \emph{semantic} conflicts.\footnote{This relates two conflict terminologies: one based on the development phase in which a conflict is detected, and the other based on the language aspect that causes a conflict. We use merge conflict and textual conflict as synonyms. Build conflict refers to syntactic and static semantic conflicts. Test and production conflicts (and undetected ones) are referred as behavioral semantic conflicts. For brevity, hereafter we omit the ``behavioral'' term in spite of focusing only on behavioral semantic conflicts in the paper.}
To make matters worse, we expect semantic conflicts to cost %\tb{I think talking about long-term costs of not detecting and resolving semantic conflicts is good and motivating; we should reserve the notion of cost for this meaning; not talk about costs of resolving conflicts, which can be confusing}\pb{Not sure what to do here and still keep consistency, as we start talking about cost in the first paragraph.} 
 more than \emph{merge} conflicts, as they are often harder to detect and resolve, and might end up negatively affecting users.

Resolving \emph{merge} conflicts is often simpler, because it mostly involves reconciling incompatible independent \emph{textual} changes in the same area of a file.
Semantic conflicts are harder to detect and fix, especially when resolution occurs long after conflict introduction, because resolving them requires handling \emph{behavioral semantic} incompatibilities--- as when the changes made by one developer affect a state element that is accessed by code contributed by another developer, who assumed a state invariant that no longer holds after merging. 
In such cases, textual integration is automatically performed generating a merged program, a build is created with success for this program, but its execution leads  to unexpected behavior caused by unplanned \emph{interference} between the developers' changes--- the behavior of the integrated changes does not preserve the intended behavior of the individual changes.
Horwitz et al.(\citeyear{integrating-noninterfering-versionsHorwitzPrinsReps1989}) put this more formally: two contributions (sets of changes) to a base program \emph{semantically conflict}--- that is, interfere in an unplanned way--- when the specifications they are individually supposed to satisfy are not jointly satisfied by the program that integrates them.

\looseness=-1
To help reduce the costs associated with semantic conflicts, we need merge tools that are able to detect them, going beyond textual line-based merge tools currently used in practice~\citep{diff3-formalizationKhannaKunalPierce2007}. 
Previous work~\citep{integrating-noninterfering-versionsHorwitzPrinsReps1989, semantic-merge-verificationSousaDillingLahiri2018} proposes semantic merge tools that rely on static analysis and model checking for detecting conflicts. 
Our previous study~\citep{da-silva2020detecting} proposes and assesses the use of \emph{unit test generation} to reveal interference.
The initial results bring evidence of the potential of using tests to detect semantic conflicts, but also show a number of limitations, including a significant  false-negative rate.

To address these limitations, we extend our previous work by proposing and evaluating new techniques (testability and serialization transformations) and integrating them into SAM (SemAntic Merge), a semantic merge tool for Java that automatically generates unit tests and use them as partial specifications of the changes to be merged, with the aim of detecting semantic conflicts. 
SAM first applies a textual merge tool to integrate the changes.
In case no textual conflicts are reported, SAM builds the four program versions associated with a merge scenario--- a quadruple $(\baseCommit END, \leftCommit END, \rightCommit END, \mergeCommit END)$ formed by a merge commit (\mergeCommit END), its parents (\leftCommit END and \rightCommit END), and a \baseCommit END commit--- optionally applying source code transformations that might increase program \emph{testability} and feed the test generation tools with objects \emph{serialized} during the execution of existing project tests. 
%These transformations are required as the changes performed during the merge scenario might not be directly accessed by the unit test tools (e.g., \texttt{private} class elements). 
%These transformations are changes directly applied to the original source code aiming to increase its testability during the generation process of test suites.
%In contrast, the code holding the semantic conflict, changed by the parent commits, might involve complex objects, which are hard for the unit test tools to deal with. 
%Aiming to support the tools, we propose using serialization when creating these complex required objects. 
Then, SAM applies four test generation tools: EvoSuite and Differential Evosuite~\citep{finding-real-faultsAlmasiHemmatiFraserArcuri2017,tutorial-using-evosuiteFraser2018}, Randoop~\citep{feedback-directed-randomPachecoShuvenduErnstBall2007}, and Randoop Clean, an adapted version of Randoop we propose here.
SAM then runs the generated tests against the four program builds, collects test failure information, interprets that with our interference criteria heuristics,  and finally reports detected conflicts.

\looseness=-1
To evaluate our tool, we perform an empirical study with a dataset of 85 changes' pairs from 51 software merge scenarios that integrate changes to the same method, constructor, or field declaration. 
%\wm{very long sentence. Also, doesn't make sense after the comma. Sounds unfinished. Good idea to split after the last ---} 
These scenarios come from open-source Java projects, and are either mined by our scripts or used in previous studies~\citep{semistructured-vs-structured-mergeCavalcantiBorbaSeibtApel2019,semantic-merge-verificationSousaDillingLahiri2018,using-information-flowBarrosFilho-2017,da-silva2020detecting}.
For each merge scenario, we invoke the mentioned unit test generation tools and check their effectiveness in detecting interference following our \emph{test-based} criteria; strictly checking for semantic conflicts would require access to the specifications of the changes or knowledge about developers' intentions. 
%For a subsample, we compare the results we obtain with the model checking based semantic merge tool SafeMerge~\cite{semantic-merge-verificationSousaDillingLahiri2018}.
Since SAM invokes the unit test tools on different versions of executables (original, transformed, and serialized), we are able to measure the effect of adopting the testability transformations and serialization techniques.
For the scenarios that SAM fails to detect an existing interference, we manually analyze the causes of the failure.
This sheds light on how SAM and the underlying unit test generation tools could be improved.
Besides that, since some conflicts might be challenging to detect, we assess whether the generated test suites can detect general behavior changes and related metrics. 
This way, we may evaluate how close the tools are to detect interference considering some behavior changes might involve conflicting contributions from merge scenarios. 

\looseness=-1
Our results show that SAM best performs when combining only the tests generated by Differential EvoSuite and EvoSuite, and using the proposed Testability Transformations (nine detected conflicts out of 28).
These results reinforce our previous findings of the potential of using test-case generation to detect semantic conflicts as a method that is versatile and requires only limited deployment effort in practice, with no need for explicit behavior specifications. 
Despite the low rate of true positives, the generated tests lead to only a few cases of false positives.
This suggests that semantic merge tools based on unit test generation, as we propose here, can help developers detect semantic conflicts early, avoiding them to otherwise reach end users as failures.
%The associated benefits are likely achieved with small false-positive costs. 
However, with the current capacity of the test generation tools, developers cannot rely solely on such semantic merge tools for detecting conflicts.
%, and as such, SAM can be considered a step closer to detect conflicts supporting developers during code integration.

%The transformations improved testability in three of the nine detected interference cases, suggesting that they might be useful for interference detection. %\pb{We should keep a note to have info on the effect of transformations on scenarios without interference, or even scenarios with interference that were not able to generate test cases, satisfying the criteria or not.} 
Our manual analysis of generated test suites lead to the identification of  shortcomings of the existing tools. 
In line with those shortcomings, we suggest three potential improvements, that involve creating relevant objects required for the declarations holding the conflict, and relevant assertions exploring the \textit{propagated} interference.
For some false-negative cases, we identify and categorize improvements that could benefit unit test generation tools.
Regarding the detection of behavior changes between commit pairs, though EvoSuite is the most successful tool detecting 53\% of all reported changes, there is no combination of tools that detects all reported behavior changes. 
As a final contribution, we provide our study sample as a dataset of merge scenarios with source code, working executables (which are necessary for running tests), and interference ground truth.
This can be used to run new studies with less effort and to replicate ours.

This study is an extension of our previous work~\citep{da-silva2020detecting}.
After our initial results regarding the detection of semantic conflicts using unit test tools, we focus on evaluating the effectiveness of our technique combined with different improvements, like the generation of complex objects and the test generation process based on a target method.
First, we propose and evaluate SAM, our semantic conflict tool based on unit test generation. 
Second, we consider a larger sample of 85 changes. 
Third, we propose and evaluate the use of serialization regarding the generation and use of required complex objects. 
Fourth, we present new criteria for detecting semantic conflicts, jointly comparing the unit test outputs of the four program versions in a merge scenario, instead of considering only three as in previous criteria.
Fifth, we use additional metrics to assess how close our tool is to detect conflicts when false negatives are reported, while we also propose and evaluate Randoop Clean, a modified version of Randoop based on the limitations reported by our first work.
Finally, we provide the dataset and scripts used to run this study~\citep{appendix-paper2}, supporting replications and new experiments.

\section{Motivating Example}\label{sec:motivating}
\noindent
%\tb{picky comment: sometimes we suffix method names with (), sometimes not throughout the paper}
To illustrate the notion of behavioral semantic conflict we explore in this paper, consider the example in Figure~\ref{fig:semantic-conflict}.
\rev{Each change in this merge scenario independently aims to eliminate a  redundancy in the \texttt{cleanText()} method, namely the two calls to \texttt{normalizeWhitespace()} whenever \texttt{cleanText()} is executed.
The illustrated class \texttt{Text} results from a merge that integrates the deletion change in green \rev{(Line~8, say from a revision \leftCommit END)} with the deletion change in red (\rev{Line~13}, say from a revision \rightCommit END).}
 This example is inspired by a \mergeCommit END commit from the project Jsoup.\footnote{\href{https://github.com/jhy/jsoup/commit/a44e18aa3c1fcd25a68a5965f9490d8f7d026509}{https://github.com/jhy/jsoup/commit/a44e18a}}
The other code lines originate from a \baseCommit END revision, that is, the most recent common ancestor of \leftCommit END and \rightCommit END.\footnote{For simplicity, we assume a single most recent common ancestor. With so-called criss-cross merge situations in git, there could be more than one.} 
\rev{As the source code in the range of lines 9 and 12 separates the two changes to be integrated, there is no textual merge conflict in this case, and we cleanly obtain the syntactically valid class in the figure.}
We can then compile, build, and execute it. %code involving such class.

\looseness=-1
\rev{The primary purpose of the \texttt{cleanText()} method is to apply some string cleaning.
% through side effects.\footnote{In this context, side effects stand for any element that can be used to access the changes applied when the target method is called.} doesn't seem to be right
For that, it calls additional methods to remove duplicated words (Line~6), comments (Line~7), and normalize whitespace (Line~8). 
These calls were added in previous changes when developers independently added calls to \texttt{normalizeWhitespace()} causing the redundancy we just discussed.
%As a result, when these revisions are integrated, the developers observed redundant and unnecessary calls to the discussed method. 
Someone may argue that these redundant calls could have been avoided by establishing a good communication channel between the involved developers, but that is often not in place.} 
%However, considering the current collaborative software development, when developers can be in different countries and time zones, and the assigned tasks have different goals and no intersection, it is plausible to understand that there is no active communication.}

\rev{Aiming to eliminate the redundancy, developers decide to eliminate one of the calls, but they unluckily do not pick the same call. 
While \leftCommit END removes the method call in Line~8, \rightCommit END removes the call in Line~13 (see \cref{fig:semantic-conflict}). 
As a result, after integrating these revisions, there is no call left to \texttt{normalize\-White\-space()}}, characterizing an undesired \emph{interference} between the \leftCommit END and \rightCommit END revisions.
This way, we might assume the occurrence of an infection, as the state of the target program is incorrect~\citep{fraser2008reachability}.

\begin{figure}[t!]
	\centering
\begin{lstlisting}[language=Java,numbers=left,escapechar=`]
  class Text {
	
    public String text;
		
    void cleanText() {
      this.removeDuplicatedWords();
      this.removeComments();
-     `{\setlength{\fboxsep}{1pt}\colorbox{excerptgreen}{this.normalizeWhitespace(); }`
    } 

    void removeDuplicatedWords() {
        ...
-     `{\setlength{\fboxsep}{1pt}\colorbox{excerptred}{this.normalizeWhitespace(); }`
    }

    void normalizeWhitespace() {
        ...
    }

  }
\end{lstlisting}
	\caption{A merge of two changes (each parent removed one of the highlighted lines) that are semantically conflicting}
	\label{fig:semantic-conflict}
\end{figure}

\looseness=-1
To detect the just discussed conflict, different approaches can be adopted, like careful code review practices and strong test suites.
However, most semantic conflicts might escape to end users.
In our example, we would have to investigate whether the defect is in the individual implementations of \leftCommit END and \rightCommit END, or in how one of them interferes with the other. 
This would require a non-superficial investigation that breaks the abstraction boundaries established by the declarations of the methods called in \texttt{cleanText()}. %It would not be enough to check the specification of \texttt{removeDuplicatedWords()}, but we would have to recognize that its implementation does not eliminate extra space after the deletion of a duplicated word.

\looseness=-1
To reduce this discussed difficulty and the costs associated with semantic conflict detection and resolution, %\tb{need to add the two arguments that we need lightweight techniques; and that specifications are usually not explicit, but implicit, so no techniques work that require explicit specifications, such as model checking} 
it is important to investigate to what extent unit test generation tools could help to reveal the kind of interference we illustrate here. 
The core idea we propose and assess in this paper is the use of \emph{generated tests as partial specifications of the code revisions to be integrated}--- tests then partially capture the effect of the changes in the revisions.
This is the basis of SAM, the semantic merge tool that we propose.

In our motivating example, SAM could detect the interference with a test that explores the contents of the \texttt{text} field.
% , which holds the propagation of the infected program~\cite{morell1990theory} caused by the conflicting contributions. 
For instance, suppose a regression test generation tool (such as Randoop~\citep{feedback-directed-randomPachecoShuvenduErnstBall2007} or EvoSuite~\citep{finding-real-faultsAlmasiHemmatiFraserArcuri2017,tutorial-using-evosuiteFraser2018}, which are invoked by SAM) generates the test in Figure~\ref{fig:test-case} when given the revision \leftCommit END as input. 
\begin{figure}[t!]
	\centering
\begin{lstlisting}[language=Java,numbers=left,escapechar=`]
class TextTestSuite {

  public void test1() throws Throwable {
    Text t = new Text();
    t.text = "the`\textvisiblespace`the`\textvisiblespace\ \textvisiblespace`dog";
    t.cleanText();
    assertTrue(t.noDuplicateWhiteSpace());
  }
}
\end{lstlisting}
	\caption{A test case that reveals the interference in Figure~\ref{fig:semantic-conflict}}
	\label{fig:test-case}
\end{figure}
%
% Maybe text.equals("the`\textvisiblespace`the`\textvisiblespace`dog") would be more realistic
%
\rev{That test passes when executed against revision \leftCommit END, which leads to a single call to \texttt{normalize\-White\-space()} when executing \texttt{removeDuplicatedWords()} (Line~6 in \texttt{Text} class).}
%At that time, the parent revision \rightCommit END has not removed his duplicated method call yet (Line~13).}
With the input illustrated in \texttt{test1}, when reaching Line~7, \texttt{t.text} stores a string similar to the test input string in Line~5 but not having the extra space character right before \texttt{dog}.
Consequently, the assert successfully evaluates.
\rev{Executing this test case against revision \rightCommit END, the test also passes as there is a single call to \texttt{normalizeWhitespace()} (Line~8).
%; at that time, the revision \leftCommit END has not removed her duplicated method call yet. 
Finally, the same test case also passes when executed against revision \baseCommit END since it has two calls to \texttt{normalizeWhitespace()} (Lines~8 and 13). 
For this reason, passing in \baseCommit END and in both parent revisions \rightCommit END and \leftCommit END, we say that test1 partially reveals a behavior that should be preserved by both revisions.}

\rev{However note that \texttt{test1} fails when executed against revision \mergeCommit END in Figure~\ref{fig:semantic-conflict}. 
The \texttt{normalizeWhitespace()} method ends up not being called when executing \texttt{cleanText()}, as explained before.}
This way, \texttt{test1}, an original behavior expected to be preserved by both developers, is not satisfied in the merged version, revealing that the changes in \rightCommit END and \leftCommit END interfere (with respect to \baseCommit END)~\citep{binkley1995program,da-silva2020detecting}.
%\rev{If we could find a test that passes in revisions \baseCommit END, \leftCommit END, and \rightCommit END, and the same test fails in revision \mergeCommit END and vice-versa, we would similarly say that parent revisions \leftCommit END and \rightCommit END interfere with each other\cite{da-silva2020detecting}.}
%
% Can we say the same for a test that passes in B, brakes in L, and passes again in M?
%
This is essentially one of the criteria that SAM applies for automatically detecting \emph{interference} by generating and executing tests, as detailed in the rest of the paper.
Making sure that \emph{interference} actually leads to a \emph{semantic conflict} cannot be automatically checked in general because it involves understanding developers' intentions or proving that implementations satisfy specifications (in this case, specifications of the changes, which are hardly available in public repositories). 
However, for simplicity, hereafter we use both terms interchangeably and distinguish only where necessary for extra clarity.

\section{Detecting Semantic Conflicts}
\label{sec:solutions}
This section presents the solutions and techniques we propose to detect semantic conflicts. 
Initially, we motivate and introduce SAM, our semantic merge tool based on unit-test generation.
Next, we present different techniques, like the use of Testability Transformations and Serialization, that we explore aiming to enhance the potential of detecting conflicts. 
Finally, we present Randoop Clean, our extended version of Randoop.

\subsection{SAM: SemAntic Merge tool based on Unit Test Generation}
\label{sec:solution}

Detecting semantic conflicts, as motivated in Section~\ref{sec:motivating} is a complex task, not supported by current merge tools.  
Aiming to support developers in actively detecting these conflicts, we present SAM, our semantic merge tool based on unit test generation~\citep{da-silva2020detecting}.
The essence of SAM is to generate and execute tests for a given merge scenario (quadruple of \baseCommit END, \leftCommit END, \rightCommit END, and \mergeCommit END commits).
These tests are executed over the different commits of a merge scenario, and after interpreting their results, the tool reports if a semantic conflict is detected.

SAM can be called right after a successful textual merge is performed. 
With the resulting merge scenario, SAM invokes unit test generation tools to generate test suites exploring the changes that have just been textually integrated.
Next, SAM executes the generated suites in the different commits of a scenario, and analyzes the test results based on a number of heuristics, reporting conflicts accordingly.
We explain in detail our heuristics later in Section~\ref{sec:conflict-criteria-heuristic}.
If a conflict is detected, the tool warns developers about its occurrence, informing the class and methods involved in the conflict.
%Figure~\ref{fig:sam} presents an overview of how SAM works.
Next, we present in detail how the Java-Maven version of SAM works and its different steps (see the SAM's workflow area in Figure~\ref{fig:approach-methodology}).

\subsubsection{Starting Point}
\label{sec:first-point}
SAM is called when a merge is in progress in a local git client.\footnote{Currently, the tool is invoked when the \texttt{post-merge} \emph{hook} is actived; that is a Git feature to execute custom scripts. This hook is responsible for performing additional and specific checks after a successful merge commit is created.}
If no merge (textual) conflict occurs, SAM is invoked to verify the occurrence of semantic conflicts involving the parent commits' contributions. 
On the other hand, if merge conflicts are reported, SAM is not called since these conflicts would require manual fixes by the integrator, eventually leading to new changes not related to the original parent commits.
For merge scenarios classified as \emph{fast-forwards}, SAM does not take any action, leaving the default merge tool to lead the integration process.
%In these circumstances, although the merge commit presents two parent commits, only one parent commit, let us say left, has effectively changed the code under integration, while the other parent, let us say right, does not perform any change.\footnote{In this scenario, right and base commits represent the same commit.
%This way, left and merge commits hold the same program behavior.
%So there is no way the merge commit presents a different behavior compared to the left after integration.}
%For a future version of SAM, we plan to adopt a different approach by not creating a merge commit until no semantic conflict is reported. 
Once SAM is called, the first action is to collect the commits involved in the merge scenario.
For that, the tool gets the \mergeCommit END \emph{commit hash} from the head of the current branch, while the \leftCommit END, \rightCommit END and \baseCommit END \emph{commit hashes} are taken by calling further git commands.
%\footnote{\texttt{git cat-file -p commit\_id}, for left and right commits, and \texttt{git merge-base left-hash right-hash} for base commit.} 
After collecting these merge scenario information, the tool advances to the next step, when the parent commits' contributions are explored and mined.

\subsubsection{Selecting Mutual Changes on Same Class Elements}
\label{sec:changes-same-elements}
In this step, the tool explores the changes performed by the parent commits, aiming to collect mutually changed elements (methods, constructors, or field declarations).
%We opt to adopt this approach considering the chances of developers interfering with each other are higher when changes are applied to the same piece of code.
We assume that the chances one developer's contributions unexpectedly affect others might be higher under such situations, as these changes might modify, for example, the same variables, changing a method behavior in undesired ways.
For that, using DiffJ\footnote{\url{https://github.com/jpace/diffj}}, SAM collects the set of Java class elements changed by each parent commit. 
If at least one element is changed in both parents, the tool moves to the next step. 
Otherwise, the textual merge operation continues without applying any further steps in its usual workflow.

%On the other hand, in case a conflict occurs when dependent methods are changed in parallel, SAM is currently missing these cases.
%However, we believe SAM would still work in cases like this, considering that the generation of test suites might be performed based on the classes holding the target changed and impacted method.
%This way, SAM might explore a new approach to determine and identify dependent methods changed and impacted in parallel.

\subsubsection{Generating Executable Files}
\label{sec:generating-executable}
%In this step, SAM focuses on getting the required inputs in order to generate the test suites in future steps. 

Since SAM invokes unit test generation tools that rely on test execution as part of the generation process, it must feed such tools with executable versions of the target code we want to assess.\footnote{A JAR file with all compiled classes of the system and the required external dependencies.}
For that, SAM generates one executable version of the system for each commit of the merge scenario (\baseCommit END, \leftCommit END, \rightCommit END, and \mergeCommit END \emph{commits}).
Executables of all versions of the system are required because SAM detects conflicts by comparing the results of the tests when executed against the different versions. 
%Later, in this section, we explain our heuristic to detect conflicts.
%\revThesis{Although a conflict occurs due to changes on the same class element, when we generate these executables, we generation of test suites might focus on }

SAM performs a sequence of checkouts for each commit hash to generate the associated executables. 
%Initially, SAM locates and instruments the \emph{pom.xml} file of a given commit, adding the plugin \emph{Maven Assembly Plugin},\footnote{\url{https://maven.apache.org/plugins/maven-assembly-plugin/}} which holds the required information to generate the executable with all related dependencies.
%This plugin presents different options to describe the future generated binary file; however, since SAM only needs the source code and associated dependencies, a default configuration is used in this context.
This way, for each target commit, SAM calls the build manager to compile the code, resulting in a JAR file created and released on the \emph{target} directory of the project.
%As a final step, SAM collects the generated file, while discarding all previous changes applied to the \emph{pom.xml} file.
Aiming to increase the testability of the target code under analysis, SAM adopts extra techniques like the adoption of Testability and Serialization Transformations, which refine the just described build process. 
%These transformations are changes directly applied in the original source code, while the adoption of serialization is based on the generation of serialized objects resulting from executing the original project test suite.
But we only explain these two techniques in Sections~\ref{sec:testability-transformations-solution} and~\ref{sec:serialization-solution}, respectively.
%We explain in detail these two approaches in Section~\ref{sec:buildtest} when we present our empirical study.

\subsubsection{Generating and Executing Test Suites}
\label{sec:generating-running-tests}
After collecting the mutually changed elements and the associated executables as explained in the two previous sections, SAM generates and executes test suites. 

The generation is driven by the classes that contain the mutually changed elements.
This way, SAM invokes each unit test generation tool once for each parent commit (\leftCommit END and \rightCommit END), as these commits are responsible for introducing the changes that could be conflicting in the \mergeCommit END commit.
If methods or constructors are mutually changed, SAM might further drive the generation of test suites based on these elements, aiming to directly explore the code where conflicts likely take place.
%REVER LINK PARA SEÇAO
SAM can be configured to work with different unit test tools, and invoke a number of them as needed. In our experiments, we use two versions of Evosuite and two versions of Randoop, as detailed later. 
One of the versions of Randoop, called \emph{Randoop Clean}, we designed with the aim of generating tests more focused on our goal of detecting conflicts; Section~\ref{sec:randoop-clean} presents our tool highlighting its changes compared to vanila Randoop.

Finally, after the test suites are generated, SAM executes each one against each of the four executables associated with a merge scenario, and collects the test results for further analysis.

\subsubsection{Conflict Detection Based on Test Results Heuristics}
\label{sec:conflict-criteria-heuristic}
This step is responsible for reporting conflicts based on the results of executing the generated tests against the four executable versions of a merge scenario. 
SAM basically checks whether any test case satisfies one of our conflict criteria, which we present next, with their motivation. They all rely on the notion of partial specification, that is, a specification that constrains behavior only for a subset of the possible inputs. 
As such, each test case is seen as a partial behavior specification, and we can then refer to the definition of interference that relies on preserving parent specifications in the merged version of the code, as discussed in Section~\ref{sec:introduction}.

%Figure~\ref{fig:conflict-criteria} represents the four conflict criteria we consider for our tool; additional discussion about our criteria can be found in Chapter~\ref{chp:background}.

%\subsubsection*{Conflict Detection Criteria}
%\label{sec:conflict-detection-criteria}
%As we discuss in Chapter~\ref{chp:background}, current merge tools can not support developers on the detection of semantic conflicts.
%Aiming to address this gap, we propose here the detection of these conflicts based on unit test generation.
%\revThesis{In this section, we present the criteria we propose to detect conflicts.
%For that, we explore test cases statuses from executions on different commits of a merge scenario and compare them with each other in order to detect a conflict.}
To detect conflicts, we rely on specific conflict criteria implemented by SAM.
%Figure~\ref{fig:conflict-criteria} represents the four conflict criteria we consider for the detection of semantic conflicts.
%Overall, the first two criteria look for test cases that present a specific behavior on the parent commits, for example, passing on the left commit. 
%In contrast, the same test shows the opposite behavior in base and merge commits, like failing and vice-versa.
%The last two criteria look for test cases that present the same behavior on the base, and both parents' commits, like failing. 
%In contrast, the merge commit presents the opposite behavior, like passing and vice-versa.
We believe these criteria cover common situations of semantic conflicts.
The first two criteria (one for each parent, i.e., \leftCommit END and \rightCommit END) seek for test cases that present the same outputs in the \baseCommit END and \mergeCommit END executable versions, but a different one in the associated \emph{parent} version.
For example, consider a test case \texttt{test1} that passes when executed against the \leftCommit END version, but fails against the \baseCommit END version. 
So we might say that \texttt{test1} partially captures the intention of the \leftCommit END change; we can then see \texttt{test1} as a partial specification of the changes of \leftCommit END.
Now if \texttt{test1} fails when executed against the \mergeCommit END version, we conclude that \texttt{test1}, the partial specification of \leftCommit END, is not satisfied in the merged version, revealing that the changes carried on by \rightCommit END interfere with the changes of \leftCommit END (with respect to the \baseCommit END commit).
So when SAM finds a test that satisfies the just mentioned criteria, it reports a \emph{(test) conflict}.
%
% Can we say the same for a test that passes in B, brakes in L, and passes again in M?
%
%This is essentially the criteria we apply for automatically detecting \emph{interference} by generating and executing tests in the rest of the paper.
%Making sure that \emph{interference} actually leads to a \emph{semantic conflict} cannot be automatically checked in general because it involves understanding developers' intentions or proving that implementations satisfy specifications (in this case, specifications of the changes, which are hardly available in public repositories).    

Now, consider a different scenario (similar to the one in Section~\ref{sec:motivating}) and criteria, where a test case \texttt{test2} passes in the \baseCommit END, \leftCommit END, and \rightCommit END versions.
We can then consider that \texttt{test2} partially captures a behavior that is  preserved by both \leftCommit END and \rightCommit END changes, and therefore we expect such behavior to be preserved in the merge too.
So if \texttt{test2} fails in the \mergeCommit END version, we conclude that a behavior that was expected to be preserved is actually not preserved, revealing that the changes in the parent commits interfere with each other (with respect to \baseCommit END). 
So when SAM finds a test that passes in the \baseCommit END, \leftCommit END, and \rightCommit END versions, but fails in \mergeCommit END, it reports a \emph{(test) conflict}.

Since our conflict criteria rely on the final statuses of test cases executed in different executable versions, we must further comment about test statuses different than \emph{passed} and \emph{fail}.
For test cases that present \emph{error} statuses, we opt to not consider them when reporting conflicts.
We believe that considering these cases might introduce false positives in our results as we could not correctly verify the test case statuses.
%For example, by executing test suites in different versions of a program, if a specific version has a call to an external service and no answer is received, the execution would break due to unexpected trhown exceptions or other unexpected behavior.
%The \emph{error} status might be caused by external dependencies, and not changes performed by the parent commits of a merge scenario.
As a final remark, it is important to discuss that our conflict criteria are valid to detect conflicts if the associated test cases explore the conflicting changes integrated during a merge scenario.
Otherwise, false positives might be reported as well.

\subsubsection{Report of Semantic Conflict Occurrence}
\label{sec:report-conflict}
Once a test case satisfies one of our conflict criteria, our tool warns the developer about a potential conflict occurrence by informing the element where the conflict takes place, as also the test that reveals the conflict.
Then, the developer might evaluate whether the reported conflict represents an actual conflict or not. 
If so, she may apply changes in order to fix the conflict and change the current \mergeCommit END commit; otherwise, she skips the warning leaving the \mergeCommit END commit without applying any change.

\subsection{Testability Transformations}
\label{sec:testability-transformations-solution}
%REVER
Previous studies~\citep{silva2017analyzing} report that, due to a number of characteristics of the code under analysis, unit test generation tools might have a hard time generating tests that detect bugs. 
In our previous study~\citep{da-silva2020detecting}, we observe similar limitations for detecting conflicts.
For example, it might be harder to generate conflict revealing tests for classes with many  private members, as these cannot be directly exercised by the tests. 
Nevertheless, directly invoking such members could reveal conflicts that would be hard to reveal by tests that only invoke public members (that indirectly invoke the private ones).
To improve testability and increase the chances of detecting conflicts, we propose here three Testability Transformations that adapt the source code of the parent commits before creating the builds and feeding the generation tools with executables. 
These transformations are motivated by preliminary experiments we performed using the unit test generation tools with toy examples and a small subsample of the scenarios we consider in our evaluation.

As just motivated, the first proposed transformation replaces non-\texttt{public} access modifiers with \texttt{public} ones; it is applied to classes, methods, constructors, and fields declarations. 
By making all elements public, more elements can be called and accessed by the generated tests, possibly increasing the chances of detecting conflicts.
This, however, brings the risk of reporting false positives, as could happen when a generated test acesses an originally private member in a way that is not equivalent to the indirect acesses from the available public members. 
There is also the risk of the tools using a significant part of the generation budget for directly calling elements not involved in the conflict, as we apply the transformation to all classes; the motivation is that calling non-related elements involved in the conflict might lead to indirect object state change that contribute to conflict detection. These aspects are evaluated in the experiments we describe later.

Our second transformation adds an empty constructor to classes lacking one, as this might help to generate tests that create and exercise objects of such classes.
We observed that this could be especially useful for classes having only constructors that require complex object structures as arguments. 
Again, this transformation brings the risk of false positives, as reported conflicts might be revealed with object states that would not be reachable with the original class.
For simplicity, in case a class doesn't directly extend \texttt{Object} no empty constructor is added, as this would potentially require adding a chain of constructors to the class hierarchy. 
%This way, when the transformations are applied, the user may inform if he/she wants to fully or partially apply them.

Finally, our third transformation handles scenarios in which the mutually changed declarations occur inside inner classes. 
As the generation tools cannot directly exercise inner classes, we extract them to the outer level. 
For simplicity, we manually apply this transformation, as it is not often required. The other two have been implemented and are automatically applied. 
%We present these transformations as an executable jar file, which covers the two first presented transformations, while the last transformation is manually applied. 
%This file receives the local path of the target class under analysis as input.
%A semantic merge tool could apply such transformations by adding this jar file as an external dependency. 
%We expect no major negative implications for the users of the tool, as applying these transformations is computationally not expensive compared to generating and executing tests.

\subsection{Serialization Transformation}
\label{sec:serialization-solution}
%REVER

As previously discussed, unit test generation tools might not be able to generate tests that exercise complex object structures in useful ways~\citep{da-silva2020detecting}. 
Such structures, however, might be required to reveal conflicts. 
Aiming to address this limitation, we feed unit test generation tools with concrete object graphs~\citep{elbaum2006carving}, which can then be used in the generated tests, increasing the chances of detecting conflicts.
We collect and serialize these object graphs by monitoring the execution of existing, manually created, project tests. 
The effectiveness of this technique is then directly dependent on the availability of project tests that manipulate complex objects. 
For projects with no tests, we do not use this technique.

% additional information, which might be used to generate relevant instances of the required objects.
% This way, consider we might use original project test suites by running them while generating a list of objects as XML files.
% So a test tool might parse those XML files, having access to well-formed objects, and directly using them on test suites. 

To serialize objects, we implement OSean.EX.\footnote{\url{https://github.com/spgroup/OSean.EX}} 
First, our tool instruments the target method--- the method under analysis--- by adding a call for an auxiliary method that is responsible for receiving and serializing the object currently executing the target method, and the arguments passed to this method.
OSean.EX also adds the auxiliary class to the original target project.
With this first instrumented version of the project under analysis, the tool runs the manually created project test suite for a specific amount of time, creating new unique serialized objects each time the target method is reached; eventually, duplicated objects are discarded.

When the project test suite execution is finished, our tool discards the instrumented version of the project and creates a new class that declares a number of methods, one for each previously serialized object. 
Each method simply deserializes an associated object and returns it. 
OSean.EX then adds this class to the original version of the target project, creating a second instrumented version of the project, and building it.
This version's executable can then be fed to the unit test generation tools, which are  able to create tests that call the deserialization methods and use the returned complex objects. 

Considering that a merge scenario has related commits, and that object serialization might be expensive, OSean.EX performs all the steps for a single commit, say \leftCommit END, depending on how it is invoked. 
In this case, for the \rightCommit END our tool would simply perform the last steps of adding the created class and building the extended version of \rightCommit END.  
The cost of using this technique depends on the amount of time allocated to execute the project test suites. 
%Since developers can manually set the amount of time the tool may spend running the project test suites, they may decide the approach that better fits their needs.

\subsection{Randoop Clean}
\label{sec:randoop-clean}

To increase the chances of detecting conflicts, we propose here Randoop Clean, which adapts Randoop with the aim of creating tests that more often invoke the method under analysis, and increasing the diversity of objects manipulated by the generated tests (see our online Appendix for more details~\citep{appendix-paper2}).
%Figure~\ref{fig:generation-sequences} shows how Randoop Clean works.
As we preserve most behavior of original Randoop, we highlight here only the changes implemented by our tool. 
First, before Randoop starts to generate tests, it selects all public methods and constructors from a list of classes given as input and puts them in a \emph{pool}. 
Next, Randoop creates test prefixes by randomly selecting elements from the pool and generating sequences of statements that invoke such elements. 
If a particular method is expected to be covered by the generated test cases, it can be given as input to the tool.
However, in our context, we want to go beyond that and increase in these sequences the number of calls to the method under analysis, that is, the method changed by both parents commits (target method).
In principle, this could increase the chances of conflict detection.

To explore that, Randoop Clean \emph{increases the number of calls to the target method} by reducing randomness. 
%For that, we use two variables: the list of valid generated sequences (\texttt{validSeqs}) and the list of input classes (\texttt{classes}). 
After Randoop Clean generates a particular number of method calls in a test prefix sequence, it does not randomly select the next method to be called.
Instead, it adds to the sequence a call to the target method.
In between forced calls to the target method we rely on randomness, As Randoop, as we expect that calls to other methods play an important role in modifying object states that might lead to a conflict. 
In this way, we ensure that more diverse objects states and configurations reach the target method calls.

\looseness=-1
To generate objects that are used as arguments or targets of method calls, Randoop follows the process presented earlier. 
Hence, the number of calls to the target class constructor and of generated objects is random. 
However, this might lead to reduced diversity of the object pool, and consequently less chances of detecting conflicts. 
To address this Randoop Clean tries to \emph{increase the number of calls to object creation operations (methods and constructors)}. 
%For that, we keep track of two variables: the number of \texttt{steps} and the list of target classes given as input (\texttt{classes}). 
After Randoop Clean generates a particular number of statements in a test prefix sequence, the tool adds calls to the object creation operations of the target class. 
If the target method requires different object types, Randoop Clean selects one type each time and then randomly picks a method or constructor from the \emph{pool} that returns this specific object type.

%TODO2: threats to validity deveria ser uma subsecao de results. tem como ajustar isso facil ou threats to validity se baseia/referencia algo de discussion?
%Acredito que pode sim; entretanto, trata-se de um formato utilizado? Digo, apresentar as ameaças na seção de resultados.

\begin{figure*}[ht!]
\centering
\includegraphics[width=16cm, height=3.7cm]{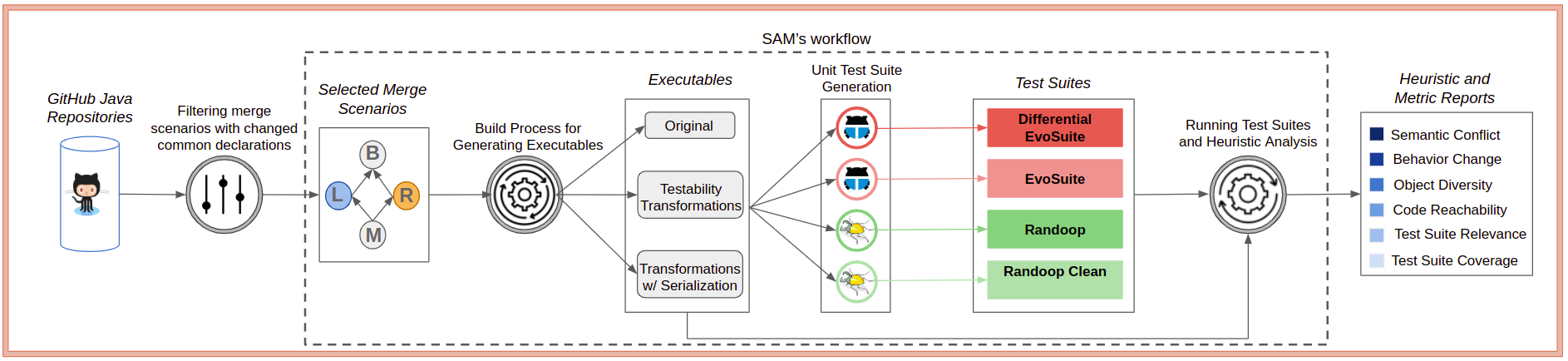} %width=13cm,height=3cm
\caption{Study setup. Starting with the selection of Java merge scenarios, we create our dataset, call the unit test generation tools, and execute the generated test suites to detect semantic conflicts. Besides that, we perform a manual analysis to explain the false positives and negatives in our sample. Inside the dashed area, we show the steps covered by SAM, our semantic merge tool.}
\label{fig:approach-methodology}
\end{figure*}

\begin{figure}[ht!]
\centering
\includegraphics[width=9cm, height=3.5cm]{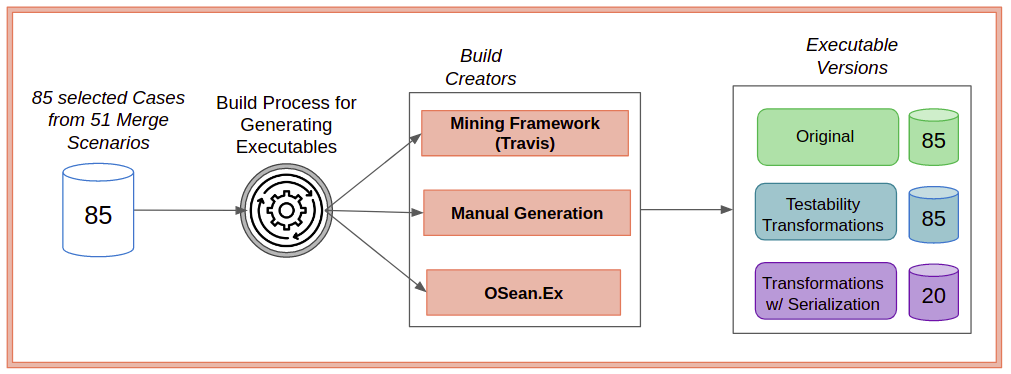} %width=13cm,height=3cm
\caption{The generation process of executables for merge scenario commits. For each merge scenario, we create a number of JAR files, which are given as inputs for the unit test tools. For the 85 cases of our sample we create JARs based on both the original code and the code resulting from applying the Testability Transformations. For a subsample of 20 cases, we create JARs based on the code resulting from applying the Testability and Serialization Transformations.}
\label{fig:executable-generation}
\end{figure}

% !TeX spellcheck = en_GB
\section{Evaluation Method}\label{sec:methodology}
\looseness=-1
\noindent
%\tb{update methodology; can be in bullet points or comments for now} 

\begin{comment}
\rev{
\begin{itemize}
	\item Sample: new merge scenarios mined from related work.
	\item Modified Randoop: it focuses on maximizing the number of different generated objects and calls for a target method by a generated test suite.
	\item When calling the test generation tools, we feed them with serialized objects previously created by running original project test suites and informing not only the target class, where a conflict takes place, and their dependencies.
	\item New semantic conflict detection criteria looking for a test case that PASS-PASS-PASS-FAIL on revisions \emph{B}, \emph{L}, \emph{R}, and \emph{M}, respectively, and vice-versa.
	\item Metrics aiming to answer how far we are to detect conflicts in false negatives: source code coverage, number of different generated objects and calls for the target method holding the conflict achieved by a generated test suite, and finally, the occurrence of behavior changes in commit pairs (a test that passes and fails in two consecutive commits, respectively).
\end{itemize}
}
\end{comment}

Our evaluation method comprises five main steps to assess the potential of SAM and unit test generation to detect interference (Figure~\ref{fig:approach-methodology}). 
First, we extract and select merge scenarios from Java projects hosted on GitHub, including a number of scenarios that appear in previous code integration conflict studies~\citep{semantic-merge-verificationSousaDillingLahiri2018,semistructured-vs-structured-mergeCavalcantiBorbaSeibtApel2019,using-information-flowBarrosFilho-2017}. 
Second, we create executable JAR files of the program versions in each selected scenario. 
Initially, we create JAR files using the original source code of the four software versions corresponding to the \baseCommit END, \leftCommit END, \rightCommit END, and \mergeCommit END \emph{commits} (see Figure~\ref{fig:executable-generation}). 
Next, we generate additional four JAR files (one for each commit version) but this time applying our Testability Transformations (see Section~\ref{sec:testability-transformations-solution}). 
For some cases of our sample,\footnote{Since we rely on the quality of original project test suites to generate serialized objects, we consider only a subsample.} we generate a third set of JAR files, now applying in sequence the Testability and Serialization Transformations (see Section~\ref{sec:serialization-solution}). 
% This new version offers serialized objects as input for the unit test tools; we feed the tools with such objects to increase the chances of creating more relevant tests.
Third, we apply four test generation tools to create tests for the parent commits of a merge scenario based on each kind of JAR file available (\leftCommit END and \rightCommit END from original, transformed, and serialized JAR files). 
Next, we run our scripts to execute the tests and discard invalid tests avoiding flakiness issues~\citep{luo2014empirical}.\footnote{We consider tests invalid if they present different results on different executions.}
Fourth, as a last automated step, we run our scripts to check the test-based interference criteria and additional related metrics regarding the quality of the generated tests. 
Fifth, we manually analyze each merge scenario and the obtained results to ensure that the reported interference is correct.
Furthermore, we investigate the reasons behind the generated tests not detecting interference in some of the scenarios that suffer from interference.

\subsection{Mining and Selecting Merge Scenarios}
\label{sec:mining-merge-scenarios}
Our dataset consists of 85 mutually integrated changes' pairs from 51 merge scenarios mined from 31 GitHub Java projects.
Since we analyze class elements mutually changed by both parents in a merge scenario, one single merge scenario might hold more than one case of mutually changed element.
As a result, for some merge scenarios, multiple cases of changed elements are considered in our evaluation.
%Based on the executable version we generate, 
%two subsamples; the first subsample owns 85 cases, while the second owns 20 cases. 
%We adopted two subsamples based on the executable types we generate, as previously mentioned. 
%To establish our sample, we follow two steps, as detailed below. 
We focus only on Java projects because the unit test generation tools we use are language-dependent, and some of our scripts are also test tool-dependent; the tools we use in our study primarily generate test cases for Java. 
Most related studies also focus on Java projects. 
We also limit our study to GitHub projects as it is one of the most popular sources of open-source projects, and most related studies also use GitHub.

From the 85 cases we consider in our dataset, 63 first appeared in previous studies~\citep{da-silva2020detecting,semistructured-vs-structured-mergeCavalcantiBorbaSeibtApel2019,semantic-merge-verificationSousaDillingLahiri2018,using-information-flowBarrosFilho-2017} that rely on datasets that share some scenarios and cases of mutually integrated changes' pairs.
Six cases from five merge scenarios come originally from \citep{da-silva2020detecting}; four cases with and two without interference.
From~\citep{semistructured-vs-structured-mergeCavalcantiBorbaSeibtApel2019}, we select eight original cases (from eight merge scenarios), two with and six without interference. 
From~\citep{semantic-merge-verificationSousaDillingLahiri2018}, we select 21 original cases (from 16 merge scenarios), three with and 18 without interference. 
Finally, from~\citep{using-information-flowBarrosFilho-2017}, we include 28 original cases (from 22 merge scenarios), 12 with and 16 without interference.

\begin{table}[]
\caption{Distribution of mutual changed class elements}
\label{table:distribution-sample}
\centering
\begin{tabular}{|l|rr|}
\hline
\multicolumn{1}{|c|}{\multirow{2}{*}{\textbf{Original Sample}}}                                    & \multicolumn{2}{c|}{\textbf{Selected Changed Mutual Elements}}                                       \\ \cline{2-3} 
\multicolumn{1}{|c|}{}                                                                             & \multicolumn{1}{c|}{\textbf{with interference}} & \multicolumn{1}{c|}{\textbf{without interference}} \\ \hline
Da Silva et al.(\citeyear{da-silva2020detecting})                                            & \multicolumn{1}{r|}{4}                          & 2                                                  \\ \hline
Cavalcanti et al.(\citeyear{semistructured-vs-structured-mergeCavalcantiBorbaSeibtApel2019}) & \multicolumn{1}{r|}{2}                          & 6                                                  \\ \hline
De Sousa et al.(\citeyear{semantic-merge-verificationSousaDillingLahiri2018})                & \multicolumn{1}{r|}{3}                          & 18                                                 \\ \hline
Barros Filho(\citeyear{using-information-flowBarrosFilho-2017})                                     & \multicolumn{1}{r|}{12}                         & 16                                                 \\ \hline
Current study                                                                                      & \multicolumn{1}{r|}{7}                          & 15                                                 \\ \hline
Total                                                                                              & \multicolumn{1}{r|}{28}                         & 57                                                 \\ \hline
\end{tabular}
\end{table}

The remaining 22 cases in our sample first appear in this paper; seven with interference and 15 without. 
All these cases come from seven scenarios that first appeared in previous work~\citep{semistructured-vs-structured-mergeCavalcantiBorbaSeibtApel2019,semantic-merge-verificationSousaDillingLahiri2018,using-information-flowBarrosFilho-2017} that considered only a subset of the cases in these scenarios.
With extra mining effort, we found out the remaining 22 cases we consider here.
In our online Appendix~\citep{appendix-paper2}, we provide further information and summarize all selected cases discussed here. 
Although we have not systematically targeted representativeness or even diversity~\citep{nagappan2013diversity}, we believe that our sample has a considerable degree of diversity concerning different dimensions such as project domain, size, and number of collaborators.

%The third part of our merge scenarios contains cases that we originally mine from merge scenarios from previous work .
%These studies are the same we mention when presenting the second part of our sample in the last paragraph. 
%Although we consider some scenarios of their original studies, we observe that multiple class elements are changed at the same merge scenario. 
%So additional cases might be mined from the related merge scenarios.
%Consequently, we use our Mining Framework to select new cases. 
%As a result, 22 new cases are added to our sample (seven merge scenarios), seven with and 15 without interference. 
%Thus, we have a sample of 85 interference cases from 51 merge scenarios.

\subsection{Building the Projects}

As mentioned at the beginning of this section, for each case in our sample we must create  JAR files that are used to generate and execute test suites. 
Considering we need to create build files with all project dependencies, for simplicity we initially try to use Travis (see Figure~\ref{fig:executable-generation}) to create such executables. 
%So our scripts request Travis to create the JAR files associated with the original source code and with the Testability Transformations.
%As a result, four binary files are generated for each version (original and transformed).
The main advantage of this approach is to reduce the chances of broken build processes due to local environment and configuration issues.
As we use the Travis infrastructure, in case of merge scenarios requiring different environment options, we would not have to deal with each one directly. 
Instead, we just set up a Travis configuration file and reuse it when applicable.
If Travis fails to create the builds due to no longer having access to old dependencies, no support for older Java versions, or by detecting problems when running additional analysis (like style checking) adopted by projects pipelines, we try to manually fix the problem on Travis by updating its configuration files; if that doesn't work, we locally create the builds. 
%We decide to further work on these scenarios for a list of reasons. 
%First, these scenarios satisfy our merge scenario criteria regarding parent commits changing the same class element. 
%Second, the fixes required for the reported issues are simple, though we manually apply them, and they did not change the original behavior of the target code. 
%Third, related studies previously evaluate these cases, so they are relevant cases to include in our sample.

The automated process involving Travis is only used for creating the builds for the original source code and the code resulting from the Testability Transformations.
The builds for the code resulting from the Serialization Transformation are created manually, as our serialization tool requires extra configuration for each project.
We have serialization builds only for a subsample of cases because that requires project test suites that exercise the method under analysis.
%In our study, we inform the method that potentially holds the semantic conflict as the target method for the serializer.
We opt for running the project test suites for 60 seconds as most project test suites in our sample are finished by this time; so allocating more time would not significantly improve the pool of serialized objects.
Once OSean.Ex accepts a list of commits, we invoke the tool giving as input the four commits in a merge scenario.
This way, the serialized objects are generated based on the first commit of this  list; in our study, the \mergeCommit END commit is always first. 
The remaining commits of that list reuse the serialized objects by deserializing them based on their versions. 
So for \baseCommit END and parent commits (\leftCommit END and \rightCommit END), OSean.Ex only performs the last step generating the executable files.

We also adopt the manual build creation process for Ant projects (one single case), as our automated infrastructure supports only Maven and Gradle projects. 
The process and infrastructure we use to create the builds appear in our online Appendix~\citep{appendix-paper2}.
%We opt to generate the binary file versions with serialized objects locally as our serializer is implemented to work that way, and it is a tool still under development.
%fitnesse

At this point, if we failed to create one of the builds for a case, we simply discarded the case in our experiment; five scenarios were discarded. 
At the end of this step, we have a sample composed of 51 \emph{merge scenarios} and 85 \emph{potential interference cases}, knowing that some merge scenarios contain more than one independent change on the same declaration.
For all 85 cases, we have executables (eight, two for each commit version in a case) with the original source code and Testability Transformations.
Finally, for 20 out of these 85 cases, we have additional executable files (four, one for each commit version in a case) with serialized objects.

\subsection{Generating and Executing Tests}

\begin{figure*}
\centering
\includegraphics[width=13cm, height=4cm]{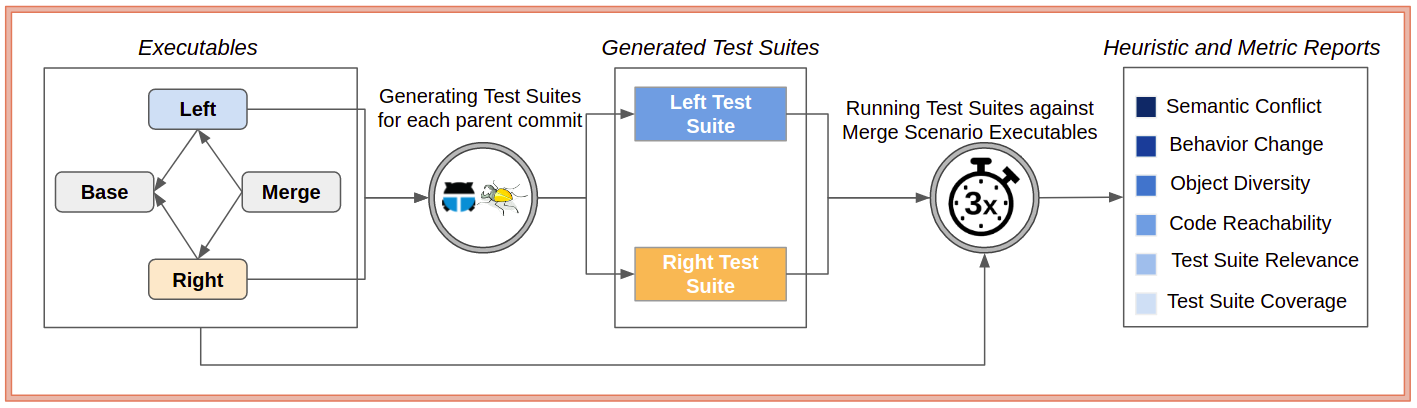} %width=13cm,height=3cm
\caption{\textmd{Generation and execution of test suites.} For each case of our sample, we generate test suites based on both merge scenario parent's commits. Next, we execute three times each generated test suite against all merge scenario commits in order to calculate our metrics.}
\label{fig:generating-executing-tests}
\end{figure*}

Each merge scenario and case resulting from the previous step has a number of proper executable files that tests can execute and exercise. 
These files are required by unit test generation tools that generate tests and run them against the system to be tested, discarding tests that fail or do not increase code coverage (see Figure~\ref{fig:generating-executing-tests}). 
This observation is valid here for the test generation tools we evaluate: EvoSuite~\citep{finding-real-faultsAlmasiHemmatiFraserArcuri2017}, Differential EvoSuite~\citep{sina2015automated}, Randoop~\citep{feedback-directed-randomPachecoShuvenduErnstBall2007} and Randoop Clean, our extended version of Randoop.
We chose the first three tools due to their robustness and popularity. 

In this step, we readily apply the unit test generation tools to create tests for four of the executable versions (\leftCommit END, \rightCommit END, and their transformed versions, as explained above) associated with each merge scenario.
For a subsample of the cases, we additionally invoke the unit test generation tools for two executable versions with serialization (serialized \leftCommit END and \rightCommit END). 
For each executable version, our scripts call EvoSuite, Randoop, and Randoop Clean passing the corresponding parent commit JAR file as input. 
For Differential EvoSuite, which tries to generate tests that reveal behavior differences between two program versions, we additionally give as input the JAR file of the \baseCommit END commit, which is used as the regression version. 
So, the tool will try to generate a test that passes in the \emph{parent} commit and fails in the \baseCommit END commit. 

For each tool, we use a budget of 5 minutes and their default configuration.\footnote{The versions of the tools used by SAM and in our study are mentioned in our online appendix.}
We decide to adopt 5 minutes considering that related work opt for different budget configurations (1, 2, or even 10 minutes); in our previous study~\citep{da-silva2020detecting}, we opt for 2 minutes.
This time, we give more time for the tools and assess whether the budget affects the detection of conflicts.

Our scripts invoke each of the four tools for the two parents in a merge scenario, considering the three kinds of executables we create (original, testability and serialization), generating then 24 ($4 \times 2 \times 3$) test suites.
For the scenarios with no serialization executable, we generate 16 test suites.\footnote{In two cases of our sample, no test suites were generated due to environment issues (one case with and another without interference). Furthermore, for some specific versions, the tools presented errors and the generation process was interrupted, resulting in no test suite. In these cases, we do not discard the cases and simply consider that the tool does not report interference.}
The number of tests in each suite varies a lot.

%Regarding the number of generated tests by each parent commit, the tools might generate different numbers.
%For example, Differential EvoSuite might generate at most one test case, while the other tools EvoSuite, Randoop, and Randoop Clean might generate multiple test cases.
%This way, we may not ensure the number of tests as it relies on the target code given as input for tools, like complexity, direct dependencies, and other associated factors.

For each resulting test suite, our scripts execute each test case three times for each of the different versions: \baseCommit END, both \emph{parents}, and \mergeCommit END (see Figure~\ref{fig:generating-executing-tests}), resulting in 12 executions. 
We execute the tests in both parents because two of our conflict criteria assess the test results against all executable files associated with the merge scenario commits. 
Finally, for each merge scenario without serialization executables, the 16 generated test suites are executed 12 times resulting in 192 executions; in the case of a scenario with serialization executables, the other eight test suites are also executed 12 times resulting in additional 96 executions. 

We execute each test case three times aiming to detect test flakiness. 
If a test case does not yield the same result (pass or break) in the three repeated runs, we filter it out, as they would not help in conflict detection due to their flakinesses. 
The test suite execution results associated with each case of our sample are grouped into three sets: tests with failed status, tests with passed status, and tests that could not be executed because they do not even compile with the version under test. 
Such validity issues with tests might occur because the test is generated for a given revision, say \leftCommit END, but is executed in other revisions as well: \baseCommit END, \rightCommit END, and \mergeCommit END. 
If the \leftCommit END revision, for example, adds a method declaration that is called in the generated test, this test will not even compile with the \baseCommit END and \rightCommit END version. 
In the same way, tests with error statuses are not considered as failed tests. 
An errored status signals an unexpected situation during test execution, which does not involve the program behavior under test.
Such invalid tests are discarded as the last action in this step, and are not used for interference detection in the next step.

\subsection{Detecting Interference}
We group test suite executions into sets for each case of our sample based on the executable versions used to generate the tests. 
Each set contains the executions associated with the \baseCommit END, parents (\leftCommit END and \rightCommit END), and \mergeCommit END commits for the original, transformed, and serialized versions. 
Next, for each execution result set, our scripts compute the test cases that satisfy one of our conflict criteria (see Section~\ref{sec:conflict-criteria-heuristic} and our online Appendix~\citep{appendix-paper2} for further information about our criteria). 
%For example, for our first two criteria, we look for tests that present the same result when executed on the \baseCommit END and \mergeCommit END commits but different results on the parent commit (\leftCommit END or \rightCommit END) (see our online Appendix~\cite{appendix-paper2} for further information about our criteria). 
%For our last two criteria, we look for tests that present the same result when executed on the \baseCommit END and parent commits (\leftCommit END and \rightCommit END) but different results on the \mergeCommit END commit. 
Finally, our scripts collect the results for further analysis and report interference if at least one test case satisfies at least one of our criteria.

\subsection{Assessing other Metrics}
\label{sec:improvements-semantic-conflict}
The steps so far are the essence for assessing the potential of SAM and unit test generation to detect interference.
However, to better understand such potential and how it is limited by the unit test generation tools we use, we go further and assess other metrics. 
In particular, in the following, we consider metrics that may help us to better evaluate the effect and limitation of each technique we rely on: conflict detection criteria, testability and serialization transformations, and unit test tools. 

\subsubsection{Behavior Change Detection}
Besides assessing whether the tests generated by the tools can detect interference, we assess whether they can establish a weaker property: behavior change between the commits in a merge scenario. 
Note that detecting interference requires detecting two behavior changes, for instance, one from \baseCommit END to \leftCommit END and another from \leftCommit END to \mergeCommit END.
So when SAM fails to detect interference we want to assess whether the generated tests could detect one or none of the behavior changes as a measure of how far the tool was to detecting interference.
To detect those behavior changes, we use the test suites previously generated and look for test cases that report different outcomes when running them against two commits. 
As we want to detect the behavior changes introduced by each parent in a merge scenario, we look for behavior involving the related parent and \baseCommit END or \mergeCommit END commits. 
So for each case in our sample, there are at most four possible behavior changes.
We then compute and compare the number of behavior changes detected by each unit test generation tool.
%\subsubsection{Test Suite Relevance}
%Considering that a parent commit might introduce multiple changes to different parts of the code, there is no guarantee that unit test tools will detect all these behavior changes due to their randomness. 
%However, we believe that the higher the number of generated tests detecting behavior changes, the higher the chances these detections are associated with different changes. 
%This way, our scripts aim to assess the number of test cases generated by each tool that detects a behavior change.
%Unlike our previous metric, here, our scripts check the number of different tests that detected any behavior change.

\subsubsection{Object Diversity and Target Code Reachability}
To compare whether Randoop Clean is closer to detect interference than Randoop, we compute two metrics that are used to compare the test suites generated by both tools:
%that emphasize the improvements implemented on Randoop Clean.
%As we advocate in Section~\ref{sec:randoop-clean}, Randoop Clean aims to maximize the number of calls to the target method holding the semantic conflict and the number of object diversity by maximizing the number of calls to create objects.
%Based on these properties, we compute two metrics: 
the number of calls performed to a target method, and the number of different handled objects. 
To compute these metrics, we instrumented both tools to collect such information and report it after the generation of each test suite.
So when we refer to Randoop in our study we actually mean a version of Randoop instrumented with this metric collection functionality.
This way, our scripts have access to, for each tool, (i) the number of calls made for all possible methods of the target classes, and (ii) the number of different objects handled by the test suite. 

\subsubsection{Source Code Coverage}
To further evaluate the improvements of Randoop Clean over Randoop, we compute the source code coverage (line, branch and instruction) of the test suites generated by each tool.
This might help understanding which tool is closer to detecting interference. 
For that, we collect only the coverage achieved by each tool against the \mergeCommit END commit. 
As the \mergeCommit END commits contain the potentially interfering changes, if these are covered by a test suite on the \mergeCommit END commit, they are likely covered on the \emph{parents} and \baseCommit END commits. 
As Randoop does not provide source code coverage information, we implement additional scripts to compute this using Jacoco~\citep{jacoco}.
For each generated test suite, our scripts call Jacoco to instrument the JAR file previously used to generate the test suite. 
Next, the test suite is executed against that new instrumented JAR file resulting in a file with the coverage results. 
Since we want to explore the coverage of a target method, our scripts compare the percentage achieved by each suite.
\subsection{Manual Analyses}
\label{sec:ground-truth}
In our study, we carry on two main manual analyses. 
First, we manually analyze each scenario in our dataset to establish interference ground truth. 
Second, after executing the experiment and detecting false positives and false negatives, we manually analyze each case and the generated tests and metrics to understand what caused the false result.

\subsubsection{Ground Truth Analysis}

Six researchers manually analyzed all cases of our sample; in pairs, the researchers individually analyzed each case to check for interference and later compared the analysis with his partner. 
If both researchers agree with the same decision, they present the scenario and its evaluation to the remaining four researchers. 
However, in case of a conflicting decision, the whole group discusses the case and reaches a verdict. 

To reduce the chances of human error and misjudgment in this process, for each interference verdict, we manually designed a test case that reveals the interference. 
Similarly, each non-interference verdict has an explanation of why we could not design such a test case; for example, one of the changes is a structural refactoring, not affecting the behavior of the other integrated changes.

Our manual analysis is local, in the sense that it involves only the mutually changed program element and its dependencies (methods and fields it calls and accesses, for example).
As we ignore the global context that depends on the analyzed program element, the changes could in fact globally interfere but we would not detect it. 
For example, say two developers, in the same method declaration, add assignments to disjoint fields of the same object. 
So, locally, the changes do not interfere with each other as they affect disjoint state elements that are unrelated in the local computations. 
But, if we consider the global context, say a method \texttt{computeRate()} that calls the changed method and compares the two fields in specific ways, we could have interference. 
Unit tests that exercise the context classes could still detect this interference by invoking \texttt{computeRate()}, but not by focusing only on the class that declares the changed method, as we do here.
We opt for checking local interference only for two main reasons.
First, it has the potential to detect a relevant part of interference cases.
Second, the global context can be significantly large or hard to capture, especially when the changes occur in widely reusable classes that are either part of an API or are invoked from multiple program entry points (as in microservices systems); manual analysis and test execution in such cases could be challenging. 

Instructions and guidelines used during this process are organized as a document, which is available in our online Appendix~\citep{appendix-paper2}.
For many cases (57), the ground truth is available in previous work, but we nevertheless follow the process above and compare verdicts. 
For all cases, we summarize the integrated changes to help reach verdicts and aid others interested in using our dataset for replications and further studies.

\subsubsection{False Positives and False Negatives Analysis}

Comparing the results of our experiment with our ground truth, we collect information on false positives (our interference criteria are satisfied, but there is actually no interference in the scenario) and false negatives (our interference criteria do not hold, but the scenario actually suffers from interference). 

Aiming to understand the limitations that unit test generation tools face--- in our context of exploiting the generated test cases to detect interference--- we analyze the test suites of the identified false negatives. 
Based on the test descriptions we wrote during our ground truth analysis, we try to manually change the unsuccessful generated test cases and check if they could then detect interference. 
As a result, we identify improvements that could be applied to the tools, as well as to better understand and help assess how close the tools are to generating a test case that would reveal interference.

At the end of this step, we obtain a dataset composed of merge scenarios associated with their build files (original, transformed, and serialized binary files), generated test suites, interference ground truth, and further information on the quality of each test generation tool.

For the merge scenarios reported with interference, we analyze the associated test suites to ensure that the tests explore the conflict that we find during our manual analysis. 
This analysis is essential since the testability transformations could introduce false positives to our results, as some semantically change the program behavior. 
For that, we check whether the failed test case assertions explore the side effects of the elements involved in each conflict.

\begin{figure}[t!]
\centering
\includegraphics[width=13cm,height=7.5cm]{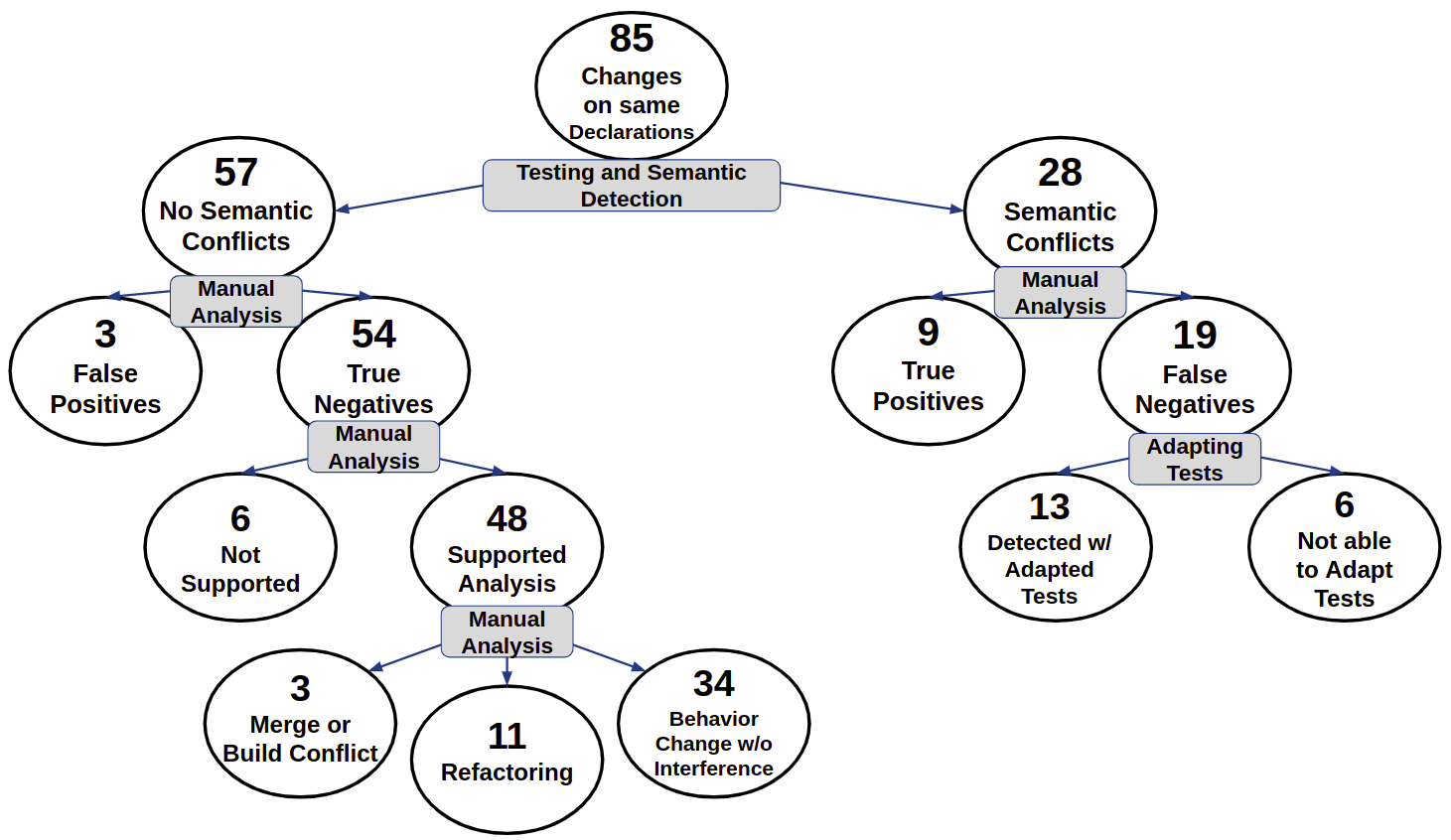}
\caption{\textmd{Conflict detection results for our dataset of changes' pairs, using the four unit test generation tools with the three kinds of executables (original, testability and serialization). Distribution of changes (on same class element declaration), their classification, and whether a conflict is detected or not.}} 
%Note: \emph{Refactor.} abbreviates \emph{Refactoring}, \emph{No Interfer.} abbreviates \emph{No Interference}. }
\label{fig:results-summary}
\end{figure}

\section{Results}\label{sec:results}
\looseness=-1 
We now present the results of our analysis of the 85 cases of changes' pairs in the 51 merge scenarios mined from 31 GitHub Java projects (see Section~\ref{sec:mining-merge-scenarios}), including how semantic conflicts are detected by SAM's interference criteria and test generation process. 
We also discuss how the test generation tools used by SAM could be improved to increase conflict detection accuracy. 
Our focus is first on the cases with conflicts. 
Later we discuss the cases with no conflicts, as defined by our ground truth, concluding with suggestions for improvement.

%\subsection{Regression Testing can be used to detect Semantic Conflicts}
\subsection{Cases With Conflicts}
%Recall that, to verify the potential of SAM for detecting semantic conflicts, we select 85 cases of changes on the same declarations in the source code from 51 merge scenarios from 31 GitHub Java projects, apply testability transformations to increase their testability, call the test-generation tools to generate test suites, and finally, execute, filter, and remove invalid test cases from our analysis. 
Figure~\ref{fig:results-summary} summarizes our results. 
The right branch focuses on the changes with semantic conflicts, while the left branch focuses on the changes without conflicts, according to our ground truth. 
Our tool could automatically detect nine out of 28 conflicts (32\%); these nine conflicts appear in five merge scenarios.
This reinforces our previous finding that unit test tools are useful to detect semantic conflicts~\citep{da-silva2020detecting}. 
Although our tool misses a significant number of conflicts, it reports only three false positives, as depicted in the left branch. 
So, based on these results, a developer using our tool should plan to use additional techniques to detect the conflicts that SAM misses, but should not worry about wasting significant time investigating false positives. 

%Analyzing the right branch in Figure~\ref{fig:results-summary}, we observe that 28 changes have semantic conflicts.\footnote{Note that, even when the scenarios stemmed from another study, we double-checked them since other works might have a slightly different understanding of semantic conflicts.} 
To illustrate one of the conflicts detected by our tool, consider the pair of changes integrated in the code of Figure~\ref{fig:semantic-conflict-real-case}, which shows a \mergeCommit END commit from the HikariCP project.\footnote{\href{https://github.com/brettwooldridge/HikariCP/commit/1bca94af9ec625f21d1b58ff10efb5be71ab87a6}{HikariCP - merge commit: 1bca94a}} 
In this merge scenario, the \leftCommit END commit adds a new condition to the if statement using the local variable \texttt{retries} (Line 6 in Figure~\ref{fig:semantic-conflict-real-case}), restricting the call to \texttt{incrementAndGet}, which increments the number of total connections to a pool. 
Independently, the \rightCommit END commit changes how \texttt{retries} is initialized (Line 3), which is referred by the changes performed by the \leftCommit END commit.
So the \leftCommit END and \rightCommit END commits individually change the program behavior (creating and adding single connections to a pool) based on their needs. 
These changes can be textually integrated with success.
No \emph{merge} conflict is reported since lines 4 and 5 separate the changes made by the two developers. 

The change from \rightCommit END, however, interferes with the change from \leftCommit END, breaking the intention of \leftCommit END's change. 
Fortunately, this is reported by our tool through the EvoSuite unit test case in Figure~\ref{fig:semantic-conflict-test-case}.
This test case was generated when our tool invoked Evosuite with the \rightCommit END serialization executable. 
As highlighted in the test case (Line 6 in Figure~\ref{fig:semantic-conflict-test-case}), the expected number of \texttt{totalConnections} is 10. 
%This assertion is defined based on the execution of the target code that returned this value during the generation process by EvoSuite.
Running this test case on the \rightCommit END commit, the test passes.
%, showing that the same behavior observed by EvoSuite during the generation process is held by the \rightCommit END commit.
For the \baseCommit END commit, the test fails as the returned number of connections is 12; as in this commit \texttt{retries} is initialized with \texttt{0}, some calls to \texttt{decrementAndGet} are bypassed (Line 16 in Figure~\ref{fig:semantic-conflict-real-case}), not decrementing the total number of connections. 
Note that on the \rightCommit END commit, the variable \texttt{retries} is initialized based on the number of acquired retries from the object \texttt{config}, which is later used in another \texttt{if} statement condition (Line 14 in Figure~\ref{fig:semantic-conflict-real-case}).
%These calls are not made as the old if condition is no longer \texttt{true} (Line 15 in Figure~\ref{fig:semantic-conflict-real-case}).
%However, these calls are made in the right version as this commit changes the local variable \texttt{retries} used in the if condition. 
For the \mergeCommit END commit, the test case also fails as the returned number of connections is -30; this time, no call to \texttt{incrementAndGet} is executed (Line 7) due to the interaction between \leftCommit END and \rightCommit END changes, which the \texttt{if} statement condition evaluate to \texttt{false}.
On the other hand, multiple \texttt{decrementAndGet()} calls (line 16) are executed.
The multiple calls for \texttt{decrementAndGet()} (line 16), in all evaluated commits, are motivated by an exception thrown due to mal-formed objects; however, the initialization of these objects is not impacted by the testability or serialization transformations presented in this study.

\begin{figure}[ht!]
\centering
\begin{lstlisting}[language=Java,numbers=left,escapechar=`]
@Test
public void test1(){
 ...
 HikariPool hikariPool = new HikariPool(hikariConfig0);
 ...
 assertEquals(`{\setlength{\fboxsep}{1pt}\colorbox{excerptgreen}{10}`, `{\setlength{\fboxsep}{1pt}\colorbox{excerptgreen}{hikariPool.getTotalConnections()}`);
}
\end{lstlisting}
\caption{\textmd{Evosuite test case that detects semantic conflict.}}
\label{fig:semantic-conflict-test-case}
\end{figure}

\begin{figure}[ht!]
\centering
\begin{lstlisting}[language=Java,numbers=left,escapechar=`]
private void addConnection(){
- `{\setlength{\fboxsep}{1pt}\colorbox{excerptred}{int retries = 0;}`
+ `{\setlength{\fboxsep}{1pt}\colorbox{excerptred}{int retries = config.getAcquireRetries();}`
  ...
  try{
+    if (`{\setlength{\fboxsep}{1pt}\colorbox{excerptgreen}{retries == 0}` && 
         totalConnections.incrementAndGet() > 
            config.getMaximumPoolSize()){
       totalConnections.decrementAndGet();
     }
     ...
  } catch (Exception e){
     ...
     if (retries++ > config.getAcquireRetries()){
        totalConnections.decrementAndGet();
     }
  }
}
\end{lstlisting}
\caption{\textmd{Test conflict caused by changes that update the same variable.}}
\label{fig:semantic-conflict-real-case}
\end{figure}

The other eight detected conflicts have common aspects with the previous example. 
First, the conflicts occur because the parent commits changes impact the values stored in the same variables or object fields. 
So, to detect the conflict, the test revealing conflict might exercise a single specific object. 
Second, the test has to directly access the object involved in the conflict (side effects of the changes), having at least one assertion that explores the changed object fields.

\subsubsection{Conflicts by Tools and Executables}
Besides understanding the potential of SAM for detecting conflicts, as reflected in Figure~\ref{fig:results-summary}, it is important to assess how each unit test generation tool invoked by SAM contributes to the overall result. 
Table~\ref{table:distribution-detected-conflicts} illustrates that, each column showing the number of conflicts detected by each tool when invoked with a specific kind of executable.

We observe that EvoSuite and Differential EvoSuite are the most successful tools, detecting six conflicts each, while Randoop and Randoop Clean detect two conflicts each. 
Most tools perform better when applied to the testability executables; only Evosuite performs better when applied to the original executables.
Differential EvoSuite and Randoop are the only tools not reporting false positives (presented in square brackets in the first line of some cells).
EvoSuite and Randoop Clean present one and two false positives, respectively.

None of the tools is able to detect all conflicts. 
Although EvoSuite and Differential EvoSuite detect the same number of conflicts, together they detect all nine conflicts detected by SAM. 
All conflicts detected by Randoop and Randoop Clean are also detected by the other tools. 
%We observe a few cases where Randoop and Randoop Clean are close to detecting the conflict, but they inappropriately explored the object holding the propagated interference. 
%For example, while exploring the object values, the related assertions only verify whether the object is null.
So to catch all detected conflicts of our sample (32\%), combining Differential EvoSuite and EvoSuite would be enough. 
A version of SAM that uses only these two tools would be computationally more efficient and detect the same conflicts as the full fledged version of SAM that invokes the four unit test generation tools. 

\begin{table}[]
\caption{Distribution of detected conflicts by unit test tools and executables.}
\label{table:distribution-detected-conflicts}
\begin{tabular}{|c|rrr|}
\hline
\multirow{2}{*}{\textbf{\begin{tabular}[c]{@{}c@{}}Unit Test \\ Tools\end{tabular}}} & \multicolumn{3}{c|}{\textbf{Executable Program Versions}}                                                                                                                                                                                                                                                                \\ \cline{2-4} 
                                                                                     & \multicolumn{1}{c|}{\textbf{Testability}}                                                                    & \multicolumn{1}{c|}{\textbf{Original}}                                                                               & \multicolumn{1}{c|}{\textbf{Serialization}}                                           \\ \hline
\textbf{\begin{tabular}[c]{@{}c@{}}Differential \\ EvoSuite\end{tabular}}            & \multicolumn{1}{r|}{\begin{tabular}[c]{@{}r@{}}\textit{6 (↓3, →3)}\\ pr. 1\\ re. 0.21\\ ac. 0.74\end{tabular}}        & \multicolumn{1}{r|}{\begin{tabular}[c]{@{}r@{}}\textit{3 (→3)}\\ pr. 1\\ re. 0.1\\ ac. 0.70\end{tabular}}                     & 0                                                                                  \\ \hline
\textbf{EvoSuite}                                                                    & \multicolumn{1}{r|}{\begin{tabular}[c]{@{}r@{}}\textit{5 (↑2, ↓4) [1]}\\ pr. 0.83\\ re. 0.17\\ ac. 0.71\end{tabular}} & \multicolumn{1}{r|}{\begin{tabular}[c]{@{}r@{}}\textit{6 (↑3, ↓5, ←1, →6)} \\ pr. 1\\ re. 0.21\\ ac. 0.74\end{tabular}} & \begin{tabular}[c]{@{}r@{}}\textit{1 (↑1, ↓1)}\\ pr. 1\\ re. 0.03\\ ac. 0.68\end{tabular} \\ \hline
\textbf{Randoop}                                                                     & \multicolumn{1}{r|}{\begin{tabular}[c]{@{}r@{}}\textit{2 (↑1, →1)} \\ pr. 1\\ re. 0.07\\ ac. 0.69\end{tabular}} & \multicolumn{1}{r|}{\begin{tabular}[c]{@{}r@{}}\textit{1 (→1)}\\ pr. 1\\ re. 0.03\\ ac. 0.68\end{tabular}}                    & 0                                                                               \\ \hline
\textbf{\begin{tabular}[c]{@{}c@{}}Randoop \\ Clean\end{tabular}}                    & \multicolumn{1}{r|}{\begin{tabular}[c]{@{}r@{}}\textit{2 (→1) [1]}\\ pr. 0.66\\ re. 0.07\\ ac. 0.68\end{tabular}}     & \multicolumn{1}{r|}{\begin{tabular}[c]{@{}r@{}}\textit{1 (→1) [1]} \\ pr. 0.66\\ re. 0.03\\ ac. 0.67\end{tabular}}                    & 0                                                                               \\ \hline
\end{tabular}
\vspace{0.2cm}

 Downward arrows (↓) stand for conflicts detected by the associated unit test tool, but not detected by the next tool below. Upward arrows (↑) stand for conflicts detected by the associated unit test tool, but not detected by the next tool above. Left arrows (←) stand for conflicts detected by the associated unit test tool, but not detected by the next left tool. Right arrows (→) stand for conflicts detected by the associated unit test tool, but not detected by the next right tool. Numbers between brackets represent false positives reported by the tools. \emph{pr.}, \emph{re.}, and \emph{ac.} stands for \emph{precision}, \emph{recall}, and \emph{accuracy}, respectively. 
 %The values of \emph{recall}, \emph{precision} and \emph{accuracy} are calculated based on the 28 conflicts with ground-truth present in our sample.
\end{table}

The results mostly show that the Testability Transformations help detecting conflicts (first column in Table~\ref{table:distribution-detected-conflicts}), but not when applied together with the Serialization Transformation (third column). 
For example, after applying testability transformations the tools could directly access objects fields and, consequently, explore them in assertions. 
So the use of such transformations leads to the detection of three additional conflicts not detected by the same tools when applied to the original executables; while Differential EvoSuite detects all three new conflicts, the remaining tools detect one conflict each.
%Comentei acima; mas as informações na tabela são apenas sobre diferenças entre as ferramentas baseado nos tipos de executáveis.
This observation reinforces our previous results that testability transformations help detect more conflicts~\citep{da-silva2020detecting}. 
Note, however, that one conflict is not detected by EvoSuite when applied to testability executable.
In this particular case, the target method is public so that the transformations would have no effect.
Regarding the addition of empty constructors, detecting some conflicts was possible after applying this transformation; in these cases, the tools focus on calling the target element instead of dealing with problems when instantiating objects.
We believe Evosuite misses this conflict in this case by chance, due to its randomness, not because of the chosen executable.
%Furthermore, the conflict is also detected when EvoSuite receives as input the binary file with serialization and testability transformations.

Regarding the use of Serialization (third column in Table~\ref{table:distribution-detected-conflicts}), eight out of the 20 cases in our subsample have conflicts.
Although the serialization numbers are quite low, remember that these numbers derive from a much smaller sample than the original and testability numbers. 
From the nine detected conflict cases with original and testability executables, only one case is in the serialization subsample. 
Moreover, this conflicting case is also detected with the serialization executable. 
Nevertheless, the detection is not due to the extra serialization information available in the executable, as the EvoSuite test case that detects the conflict (third column, second row in Table~\ref{table:distribution-detected-conflicts}) does not explore serialized objects.
So we have no evidence that the serialization technique can help improve conflict detection, but we also have no evidence that it can hinder conflict detection, as might be suggested by the illustrated number if one does not know that they are based on a subsample.
We should, though, run new studies with a bigger sample in order to better assess this issue.

Comparing our results with previous work~\citep{da-silva2020detecting} shows that our initial results are replicable; all four conflicts previously detected are also reported in the study reported here. 
We adopt the same Testability Transformations in both works, but a larger budget (5 minutes) for the test generation process. 
%So we may conclude that, at least for these four conflicts, the adopted budget does not impact the detection of conflicts. 
%Someone may argue that the budget adopted for the tools to generate the test suites negatively impacted the potential of generating more relevant tests. 
%This way, giving more time to the tools, especially EvoSuite, might detect more conflicts. 
%However, previous studies adopt a similar budget when using the unit test tools under analysis. 
%In order to evaluate the impact of higher budgets, considering the complexity of the projects of our sample, we could run a new study adopting different combinations of budgets.
%TODO2: achei isso acima meio confuso e ate contraditorio, mas posso nao ter entendido o que voce quiz dizer; podemos discutir.
%
Like our previous study, all nine conflicts detected here are detected through the same conflict criterion (see our online Appendix~\citep{appendix-paper2}); a test case that passes on the parent commit and fails on the \baseCommit END and \mergeCommit END commits. 
Although no conflict is detected using our new two criteria, they are still valid as they explore scenarios not supported by our previous criteria (see Section~\ref{sec:motivating}). 
We believe our approach for generating the test suites do not favor the new criteria. 
As we currently generate tests based on parent commits, these tests are expected to pass on these commits, partially limiting our new two conflict criteria. 
To better evaluate the new criteria, we should run a new study generating tests against the \baseCommit END and \mergeCommit END commits instead of only parent ones.

As a final remark, based on our sample, Differential EvoSuite together with EvoSuite is the best configuration of tools to be adopted by SAM. 
Similarly, we should configure SAM to use only testability executables.
Finally, regarding our proposed conflict criteria, the first criterion is the best option detecting all conflicts reported in this study.

\subsubsection{False Negatives}
Moving to the 19 cases of false negatives in the right branch of Figure~\ref{fig:results-summary}, we try to understand the limitations behind the missed conflicts. 
To this end, we manually analyze the generated test suites and check whether the conflicts could be detected after applying a few changes to the test cases. 
Our main goal is to evaluate how close the generated tests are to detecting conflicts.
To guide us during this adaptation process, we consider the test descriptions that we  created when establishing ground truth (see Section~\ref{sec:ground-truth}). 
The manually adapted test cases could detect conflicts in 13 out of 19 false negatives cases in our sample.
This suggests that improving SAM's test generation process, or allocating more resources for generation, could maybe significantly reduce the chances of false negatives. 
However, detecting the other six false negatives would likely require radical changes on how SAM works. 

To understand how close SAM gets to detecting conflicts, we illustrate one of the 13 false negatives in Figure~\ref{fig:semantic-conflict-false-negative}. 
Both parent commits add calls to methods that update the same object stored in the field \texttt{logger}, as they write different values on the log output.\footnote{\href{https://github.com/spring-projects/spring-boot/commit/fdd3f12ee0f92ac18844c08bf71df39feebb6673}{Spring Boot - merge commit: fdd3f12}} 
While the \leftCommit END commit updated the message using the method \texttt{info} (Line), the \rightCommit END commit updates using the method \texttt{debug} (Line 9). 
Although the parents use different writing methods to update the object, they are writing on the same character stream. 
To detect such interference, SAM would have to generate at least one test case with an  assertion that explores the fields of the object changed by the target method, the object stored in \texttt{logger} in this case. 
However, by manually inspecting the generated test cases, we only find one such assertion, which simply checks whether the object is null. 
Regardless of the particularities of this scenario, a conflict revealing assertion would have to compare the contents of the object stored in logger, not simply that it is different from null. 
This case shows that a generated test case reaches the interference location, the object state is infected, the interference is propagated~\cite{voas1992pie}, but the test case assertions fail to explore that. 
As such, we consider that SAM is close to detecting interference in this case, but misses it.

\begin{figure}[ht!]
\centering
\begin{lstlisting}[language=Java,numbers=left,escapechar=`]
void logAutoConfigurationReport(boolean isCrashReport) {
     ...
     if (...) {
+       `{\setlength{\fboxsep}{1pt}\colorbox{excerptred}{this.logger.info}`(" Error starting ...");
     }
     if (...) {
+       `{\setlength{\fboxsep}{1pt}\colorbox{excerptgreen}{this.logger.debug(getLogMessage(this.report));}`
     }
}
\end{lstlisting}
\caption{\textmd{False-negative involving updates to the same object.}}
\label{fig:semantic-conflict-false-negative}
\end{figure}

In other cases, SAM is not even close to detect a conflict. 
The methods holding the conflict are called by the generated test cases, but with arguments that are unable to lead test execution to reach the interference location; for example, a \texttt{null} argument makes the method raise an exception before even reaching the interference location.
The same happens when infection depends on more complex object graphs that not easily created by the test generation tools.
In these cases, with no infection and propagation, the assertions are often far from the ones that could detect the interference, making it harder to manually adapt the generated test cases.
%Consequently, irrelevant assertions are generated, which could not detect the interference even if it is propagated. 

Another major SAM limitation we observe is when infection or even reachability would only occur if the generated test cases were able to set dependencies to external resources such as database sessions. 
As this is highly project dependent, SAM would need to have access to project specific setup information, and feed them to the unit test generation tools, in order to avoid missing conflicts. 
The current version of SAM has no such support, reducing the chances of detecting conflicts  in projects that demand external resources.
In one scenario, for example, the interfering changes are made inside a block that is only executed if a valid database session is available.
%\leftCommit END commit updates an \texttt{if} condition, while the \rightCommit END commit adds another \texttt{if} statement and its related code inside an old \texttt{for} statement. 
%Since these changes are placed into a \texttt{for}, which , the new changes would hardly be reached during execution.
%It occurs because the tests do not handle external dependencies like the case just discussed here.
As no such session is set up by the generated test cases, the interference location is never reached.
%In order to address this limitation, the tools should set up the required external dependencies before executing the test cases.
By adapting the test cases with Junit  annotations like \texttt{Before} or \texttt{BeforeEach}, or using \emph{mocks}, could overcome this limitation.

%\begin{figure}[ht!]
%\centering
%\lstinputlisting[language=Java,style=excerpt,numbers=left,escapechar=`]{snippets/CloudSlangTestSuite.java}
%\caption{\textmd{Slightly adaptation of generated test case.}}
%\label{fig:test-case-cloud-slang}
%\end{figure}

\subsection{Cases Without Conflicts}
Focusing now on the left branch in Figure~\ref{fig:results-summary}, we discuss the 57 cases that have no conflict. 
We first discuss the three false positives in our sample. 
Later we discuss and classify the true negatives, as this sheds light on how SAM's dynamic analysis solution could compare to static analysis based solutions, which might conservatively err on such cases. 

\subsubsection{False Positives}
SAM might report false positives for a number of reasons.
First the generated test cases are not guaranteed to be free from flakiness. 
So, for instance, we can have a test case that fails in the \baseCommit END in some executions and passes in others.
SAM could then consider one of our interference criteria to hold when relying on an incorrect test result, and consequently wrongly report a conflict.
To reduce this problem, we run each test suite three times in each commit, aiming to detect and discard flaky tests by comparing the results of the three test executions; if the three results are not identical, we discard the test case and do not consider it for detecting interference.
However, this is not enough to eliminate flakiness in general.
In fact, in our study we still observe two (out of three) false positives due to flakiness.

The remaining case of false positive observed in this study actually corresponds to an interference, but not one caused by the changes in the analyzed method.
As our manual analysis is local, focusing only on the analyzed method and its dependencies, we conservatively classify this case as a false positive.
The parent commits change a common target method,\footnote{\href{https://github.com/spring-projects/spring-boot/commit/958a0a45f164601d01cb706c19f22ed3e25eff56}{Spring Boot - merge commit: 958a0a4}}, each setting the values of disjoint sets of object fields.   
Additionally, the \leftCommit END commit updates the version of an external dependency.
%While the \rightCommit END commit sets the values of nine fields of an object called \texttt{builder}, the \leftCommit END commit also changes some fields of the same object but disjoint ones; additionally, the \leftCommit END commit updates the version of an external dependency. 
The generated test case based on the \rightCommit END commit passes in that commit as expected (see Figure~\ref{fig:semantic-conflict-false-positive}), as its changes lead to an exception during execution (\texttt{IlegalStateException}). 
When running the same test on the \baseCommit END commit, the expected exception is not thrown and the test fails.
% as the sets applied by the \rightCommit END commit are not performed on the \baseCommit END commit (Line 7 in Figure~\ref{fig:semantic-conflict-false-positive}). 
Finally, the test fails again when run against the \mergeCommit END commit, but this time due to the new external dependency version integrated by \rightCommit END, which leads to a different unexpected exception.
% (Line 4 in Figure~\ref{fig:semantic-conflict-false-positive}). 
This way, we have a test case that satisfies one of our conflict criteria, but only part of the failed states are caused by the parent changes on the common target method. 
%If we skip the code execution that caused the failure on the \mergeCommit END commit, the expected exception would be thrown, leading the test case to a successful state, and consequently, no conflict would be reported as our conflict criterion would not be satisfied.

\begin{figure}[ht!]
\centering
\begin{lstlisting}[language=Java,numbers=left,escapechar=`]
@Test
public void test1108(){
   ...
   MongoClientOptions mongoClient = builder13.build();
   try{ 
      MongoClient mClient = mProp.createMongoClient();
      Assert.fail("Expected exception of type" +
          +"java.lang.IllegalStateException; ... ");
   } catch (`{\setlength{\fboxsep}{1pt}\colorbox{excerptgreen}{IllegalStateException}` e) {
      ...
   }
   ...
}
\end{lstlisting}
\caption{\textmd{Test case associated with false positive caused by unrelated parent conflicting contributions.}}
\label{fig:semantic-conflict-false-positive}
\end{figure}

Another reason for false positives is that our interference criteria are simply approximations, as interference is not computable in general. 
SAM uses them regardless of the characteristics of the generated tests, but the criteria is guaranteed to be valid only if the test assertions solely explore the state elements affected by the changes of the parent commit that was used to generate the corresponding test.
For example, consider a merge scenario with \leftCommit END and \rightCommit END commits that simply change how a disjoint set of state elements is updated.
Say \leftCommit END adds the assignment \texttt{left=1} and \rightCommit END adds \texttt{right=1}, whereas both variables were initialized with \texttt{0} in \baseCommit END. 
If SAM generates a test for \leftCommit END asserting \texttt{left==1 \& right==0}, this test breaks in \baseCommit END (as \texttt{left} evaluates to \texttt{0}), passes in \leftCommit END, and breaks in \mergeCommit END (as \texttt{right} evaluates to \texttt{1}). 
As this test satisfies our criteria, SAM wrongly reports interference even though the integrated changes do not affect each other.
The false positive is caused by the test assertion that unnecessarily constrains the value of the \texttt{right} variable, which is not affected by \leftCommit END's change.
We, however, observed no such cases in our sample. 

\subsubsection{True Negatives}
%Overall, SAM does not report many false positives (three cases). 
%Most cases without conflicts are correctly classified as true negatives. 

Moving now to the 54 cases of true negatives, six are classified as not supported because the potentially conflicting changes occur in test cases and test classes, not in the code that implements system functionality.
So even if the changes were conflicting, SAM would not be able to detect that as its current version does not support test classes; the unit test generation tools are not configured to generate test cases for test classes, as the associated testing framework and project environment are not fed to the tools, which would need to be adapted for such purpose.
%Even if we could provide the requested environment, we would still classify the changes without semantic conflicts based on our manual analysis and interference criteria. 
%The changes involve refactorings or semantic changes, but they are not interfering with each other. 
%In two cases, although the changed elements are also placed in test classes, the parent commits change regular methods, so the tools might generate tests exploring them.

For 34 of the 48 changes supported by SAM, the parent commits individually change  program behavior, but when integrated the changes do not locally interfere with each other. 
Static analysis tools that rely on more conservative analysis could maybe err on that, but the chances of SAM erring on those cases are reduced.
The same applies for the 11 cases in which one of the integrated branches only applies whitespace changes and structural refactorings such as renamings and extractions of variables and methods. 
In such circumstances, even if one parent commit changes program behavior, we have no interference and semantic conflict. 
This way, even if the generated tests present different results between the \baseCommit END and one parent, let us say \leftCommit END, the \mergeCommit END commit behaves exactly like the \leftCommit END, as \rightCommit END does not change program behavior. 

Finally, for the remaining three cases, the parent commit changes cause other kinds of conflicts during the integration (textual or build conflicts). 
%Our goal in this study is to analyze the interference among contributions, so it is essential to ensure the class files that hold the semantic conflicts have only the changes performed by the parents' commits. 
When merge or build conflicts occur, human intervention is necessary to fix these conflicts, and their resolution often involves discarding some of the changes. This in turn likely reduces the chances of a residual semantic conflict. 
%then it would be more difficult to isolate the parent commit changes and perform our analysis. 
%Even with this risk, 
That is what we observed in these three cases: significant parts of the changes were discarded during the textual and build conflicts resolutions, and so no conflict was wrongly detected by our tool.

%REVER Tabela...
%Regarding the accuracy of the evaluated tools, we observe that Differential EvoSuite presents the highest value for this measure (0.74, third row, second column in Table~\ref{table:distribution-detected-conflicts}); it occurs due to its precision of 1.0, though it presents a recall of 0.21. 
%Next, we have EvoSuite with an accuracy of 0.71 (fourth row, second column in Table~\ref{table:distribution-detected-conflicts}); the precision of EvoSuite (0.83), compared to Differential EvoSuite, is inferior due to the occurrence of false positives, as we previously discussed here.
%We will discuss their occurrence later in detail. 
%Finally, Randoop and Randoop Clean present similar results for accuracy (0.69 and 0.68, fifth and sixth row, second column in Table~\ref{table:distribution-detected-conflicts}, respectively) but with a small recall (0.07). 
%Although we have applied improvements on Randoop Clean, we do not change the core steps adopted by the tools when generating test suites. 
%This way, for example, the non-detection of conflicts caused by the creation of irrelevant complex objects is still a weakness in both tools.

\subsection{Further Evaluation of Tools and Related Test Suites}
Although our main focus here is the detection of semantic conflicts, we are also interested in evaluating other metrics regarding the quality of the generated test suites.
This way, in this section, we explore metrics that allow us to better understand the strengths and weaknesses of our approach and the evaluated unit test generation tools.

\subsubsection{Behavior Change Detection}
Regarding the detection of Behavior Changes, we are looking for test cases that present different outputs when executed against a pair of commits. 
This way, we might evaluate how close the tools are of detecting conflicts, in case the reported behavior change is associated with the same changed class element. 
%Additionally, the reported change must also be associated with related changes performed during the parent commits.

Overall, 89 behavior changes are detected by the generated test cases when using all kinds of executable we consider in our experiment. 
These changes involve either a parent and the \baseCommit END, or a parent and the \mergeCommit END commit. 
EvoSuite is the most successful tool detecting 47 out of 89 behavior changes. 
Next, we have Randoop Clean, Randoop, and Differential EvoSuite with 38, 37, and 28 detections, respectively. 
Even combining the last three tools, they would not achieve the rate detection of EvoSuite, as their outputs overlap.
Different from the results of conflict detection, Differential EvoSuite does not rank first this time. 
%We believe how each tool works and uses the time budget may motivate these results.
%However, a new study is required in order to compare the detection of behavior changes under different budgets.
As each tool could detect at least one exclusive behavior change, no combination of tools could capture all reported behavior changes.
However, the highest detection rate not including all tools could be obtained by combining EvoSuite, Randoop Clean, and Differential EvoSuite, as they report together 85 out of 89 behavior changes.

%Regarding the intersection of detected behavior changes by tools, we observe Randoop and Randoop Clean present the highest intersection, with 30 changes in common.
%We believe these detections are motivated by the original behavior of Randoop, which is not changed/extended by Randoop Clean.
%Next, we have EvoSuite and Differential EvoSuite detecting the same 14 changes, while all tools detect only four changes of all changes under analysis.
%This way, we may conclude that based on the particularities of the behavior changes, specific unit test tools are more prominent to detect them.

We observe that the adoption of Testability Transformations help the tools to detect behavior changes. 
In the same direction as for semantic conflict detection, 20 additional changes are detected with testability executables, when compared to the original executables (only 69 changes detected). 
From the 19 false-negative cases observed in the experiment, we could detect behavior changes in 11 of them.
However, the reported behavior changes are not caused by the changes involved in the semantic conflicts.
So we can't say that the tools were close in these cases.

We also observe that the serialization executables help detect 15 changes not covered when using the original and testability executables. 
In these cases, as the tools have access to realistic objects, the generated test cases may further and deeply explore the instructions and branches of the target code under analysis.

\subsubsection{Randoop Clean Evaluation}
We evaluate Randoop and Randoop Clean with respect to the following metrics: \emph{Target Code Reachability}, \emph{Object Diversity}, and \emph{Source Code Coverage}.
Compared to Randoop, Randoop Clean generates more tests that directly call target methods. 
Regarding the diversity of objects, no matter the kind of executable used in the experiment, Randoop Clean often generates more diverse objects than Randoop when the time budgets are not inferior to 5 minutes; the larger the budget the greater the difference in favour of Randoop Clean.
Finally, overall, the tools present similar coverage in most cases.
Likely due to our sample size and the reduced effects of the Randoop Clean changes--- its benefits cannot be observed in cases where the tools fail to reach the target method--- we observed no statistically significant difference between the tools.
Similar results are observed when using original and Testability Transformations executables. 
However, using executables with serialized objects lead to higher coverage.
Thus we may conclude that the quality of serialized objects allows both tools to more deeply explore instructions of the target code under analysis.

\section{Discussion}\label{sec:discussion}
\noindent
% The occurrence of conflicts negatively impacts team productivity and software quality. 
% Depending on the kind of conflicts, like merge conflicts, there are novel techniques to support their resolution or even avoid developers spending time fixing them. 
% Even with new ways to handle merge conflicts, these new tools do not support behavioral semantic conflict detection or resolution. 
% While no approach is available to detect these conflicts, they will continue to occur and, at some point, show up to the team or even the final user. 
% The later these conflicts are observed, the harder it will be to fix them.

Semantic merge tools based on regression testing, as we evaluate here, can help developers detect semantic conflicts. 
Due to the observed low number of false positives, the benefits can be obtained by avoiding major costs on wasted developer effort. 
However, due to the significant number of observed false negatives, developers should not exclusively rely on our semantic merge tool to detect semantic conflicts. 
They should still try to catch such conflicts by reviewing the code and executing project tests.

%First, we assess the detection of semantic conflicts based on a semantic merge tool using unit test generation. 
Although we evaluate the use of SAM with four unit test tools, our results show that combining EvoSuite and Differential EvoSuite would be the best option to detect all conflicts in our sample. 
Configuring SAM that way we might spend less time generating tests and detecting conflicts. 
%The detection of conflicts relies on heuristics based on the different program behaviors held by the merge scenario commits. 
Although we present four conflict criteria, not all of them detected conflicts in our study. 
However, we would not suggest a version of SAM that only applies a subset of the criteria, as checking them is not expensive.

% Second, in this study sample, our results report only a few false positives (three cases), which could be addressed by further exploring the failed assertions reported by the test executions. 
% For example, in case our conflict criteria are satisfied due to different failed assertions, our scripts might check whether all merge scenario commits fail/pass on the same test case assertion. 
% Additionally, if our conflict criteria are satisfied but the related test case does not exercise the target method, our scripts might double-check whether the target method is executed during the test case execution. 
% If the target method is not executed, we might assume that no conflict involving the target method is detected. 
% This way, if a semantic merge tool based on unit test generation is available for a development team, we expect developers would most of the time be warned about real conflicts---the team productivity would not be affected for no reason. 
% In the worst scenario, the tool would never warn about any conflict, which is the same scenario of no tool supporting semantic conflict detection.

Although the proposed testability transformations are not sound, in our sample we observe that the transformations contribute to increasing the testability of the code under analysis without drawbacks, allowing the tools to directly access and call all elements of a target class.

We believe the adoption of serialization may support the detection of conflicts, though we provide no such evidence in our study, likely due to the restricted subsample we consider for evaluating the serialization transformations. 
As the quality and coverage of project test suites play an essential role in providing diverse serialized objects, projects with weak test suites won't benefit from serialization.
In our evaluation, only nine conflicting merge scenarios were associated with project test suites covering the target methods where the conflicts takes place.
%As we previously explained, serialization addresses an essential weakness of unit test tools, namely the creation of complex objects. 
%While tools face problems generating these objects with multiple dependencies, for example, serialization provides authentic and relevant instances of those objects for the tools during the generation process.  
%This way, we expect the target code might not only be reached but also with relevant objects, which would lead to the eventual interference state. 
For the sample of nine scenarios with conflicts, we observe strong test suites associated with eight cases, covering the target method and providing more than 100 serialized objects for each case. 
However, these objects were not diverse enough to reach the infection state.
%As a result, different parts of the target code might be covered, as we observe in our results here.

A special benefit of our regression testing approach to detect conflicts is that one ends up with a test case that reveals a conflict, when the tool reports one. 
This decreases the effort to understand how a conflict occurs. 
The test case limits the amount of source code that should be analyzed and changed to fix the conflict, and can be used for debugging and understanding the mechanics of the conflict. 
Finally, the test case could also help making sure the conflict is fixed. 
This contrasts with static analysis approaches, which report a conflict and maybe the statements involved in the conflict mechanics, but provide no test case. 

Although our study and results are restricted to Java, we could use a similar approach for other languages with support for unit test generation tools like those we use here.

% Although our primary goal in this study is to evaluate the potential of detecting semantic conflicts using unit test generation tools, we are also interested in evaluating their potential by detecting general behavior changes.
% As previously motivated, if a behavior change is detected due to changes applied to the method holding a semantic conflict, we might conclude that the tools are half a way to detect the conflict.
% Unlike the reported test conflicts, many cases of our detected behavior changes are captured by exceptions thrown during the target code execution.
% Detecting a behavior change under these conditions represents a new pattern of detection by tools that we do not observe yet.
% This observation allows us to conclude that eventual test conflicts that might result in thrown exceptions might be captured by our current approach.
% However, in our sample, we do not have cases of conflicts under these conditions.

As a final remark, although SAM focuses on detecting semantic conflicts, it does not support fixing the conflicts due to the particularities of this conflict type.
Different from build conflicts~\citep{dasilva2022build}, test conflicts are harder to fix as developers must take into account the semantics of the integrated changes that cause the conflict. 
This way, applying straightforward changes, like those adopted for build conflict fixes, is often not enough in this context. 
As a result, to fix test conflicts, developers must be aware of the program specifications or intents, and then, they might apply the required changes aiming to meet that.

\subsection{Improving SAM}
\label{sec:improvements-sam}

Our evaluation of SAM reveals improvements that might be implemented in future versions of our tool.
We also discuss a number of usage scenarios for SAM, and different contexts in which our tool might be adopted. 
%Additionally, we present different approaches that might be used to get a good impression when adopting SAM.

\subsubsection*{Using SAM in a Reactive Way to Detect Conflicts}
Knowing that SAM requires significant computing resources to generate the test suites and execute them on each commit of a merge scenario, SAM's usage by individual developers might require support from a server that runs the analysis without blocking local repository activities. 
So a developer would merge locally and move on while the analysis is performed on the server. 
The developer is later notified.

%For example, consider that SAM is used with a budget of 5 minutes to generate a test suite, and each execution of the tests on each merge scenario commit takes 3 minutes. 
%Consider SAM is used with a budget of 5 minutes to generate a test suite, and execute each test suite three times on each commit, in the end, this step would take 41 minutes, which is a significant amount of time for a developer waiting for an answer. 
%TODO2: nao sei se entendi direito, mas essa parte acima nao parece estar correta. eh preciso informacao de quanto tempo a suite gerada leva pra executar. 5 minutos pra gerar a suite nao implica que ela leva 5 minutos pra executar.
%Furthermore, the developers should offer an environment supporting the execution of the steps previously described. 
%So such a concern is a valid issue as it might negatively impact the use of SAM.
%

Alternatively, SAM can also be integrated to continuous integration pipelines, in such a way that contributions to be integrated into a main remote project repository are first analyzed by SAM before integration actually takes place.
These are just two contrasting usage scenarios, but the tool could fit other scenarios as well. 
%If a conflict is reported, this information might be displayed on the PR page or other information source, and the associated roles can take knowledge of the problem. 
%So the developer, who sends the contribution, would not waste time waiting for an immediate answer and keep working on other tasks, while no local private environment would be required.

%TODO: SEE THE USE OF THE TERM FULL SPECIFICATION.
\looseness=-1
\subsubsection*{Reusing Original Project Test Suites as Input for SAM}
For projects with solid and robust test suites, we could use a configurable version of SAM that extends the current version in such a way that the detection of conflicts relies not only on generated tests, but also on existing and manually created project test suites. 
This combination of generated and existing project tests could increase the potential to detect conflicts.
%Instead of using unit test tools to generate test suites, SAM would reuse the project test suites, and, based on our conflict criteria, SAM would report the occurrence of conflicts.
%If a parent adds new tests during its contribution, these tests might also be used by SAM.
%Similarly, for projects that follow good software engineering practices, like the adoption of CI and committing new contributions or bug fixes with the associated test, the potential to detect conflicts is higher.
%Knowing that these new tests might explore new elements added or changed by a parent commit, running them in the \baseCommit END commit might result in error statuses.
%However, SAM already handles these issues by discarding those tests (see Section~\ref{sec:methodology}).

We also envision a version of SAM that generates test suites not only for the parent commits in a merge scenario, but also for the \baseCommit END and \mergeCommit END commits. 
If the interference criteria applies for these test suites, we could similarly report conflict.
This could also increase the potential of detecting conflicts, but further studies are necessary to assess that.
%Still about the use of \emph{full specifications}, the generation of test suites by unit test tools based on parent commits could be extended to other commits as well.
%For example, SAM might generate tests based on the \mergeCommit END commit.
%As a result, instead of using generated tests as \emph{partial specifications}, SAM would use them as \emph{full specifications}, as the target code used to generate the tests holds all changes performed by the parent commits.

% Additionally, we might consider a third way of using SAM combining the use of generated and original project test suites. 
% For example, for projects with test suites that do not detect conflicts, SAM might start the generation of new tests and further execute them to detect conflicts. 
% This way, we might offer SAM as a customized tool, as the developer might choose which option better fits the project context: (i) reusing original project test suites, (ii) generating new test suites, and (iii) combining previous options.

\subsection{Improving Unit Test Generation}
\label{sec:improvements}
We observed a few limitations and weaknesses of generated unit tests in our specific context.
%regarding how the test generation tools work and the quality of generated test suites.
For instance, in a few cases the generated tests are not able to create complex objects with internal or external dependencies. 
As presented in Section~\ref{sec:methodology}, we try to address some of these limitations by applying testability transformations in the code under analysis. 
Although this helps, a number of limitations persist.
By manually analyzing generated test suites of false-negative cases, we were able to better understand the limitations.
We discuss the main ones in the following.

%Recall that, during our manual analysis, we document the conflicts in our scenarios in detail.
%As previously explained, during our manual analysis, to check whether the merge scenarios have or not test conflicts, we create a test description for the merge scenarios we classify with conflicts.
%Based on these descriptions, we try to apply changes to the generated test suites of false negatives and verify whether the test suites could detect the conflicts with the minimum possible number of changes. 
%Interestingly, the improvements we propose here are not fully exclusive to detect test conflicts, but can help to improve unit test generation tools in general.

\subsubsection*{Reaching Interference Location through Relevant Object Creation} 
The unit test generation tools have a hard time creating tests that manipulate objects that need to be directly or indirectly exercised in order to detect a conflict. 
Many test cases prematurely finish their executions due to failed attempts to access fields or call methods on objects that are not properly initialized or configured.\footnote{This case refers to merge commit \href{https://github.com/jhy/jsoup/commit/a44e18aa3c1fcd25a68a5965f9490d8f7d026509}{a44e18a} in project Jsoup.} 
The attempt to access fields through a variable \texttt{textNode0} throws a \texttt{NullPointerException}, since this variable is not properly instantiated with a valid object. 
Giving relevant objects for test cases like this is necessary to at least reach the interference location. 
%Based on this, we extend this test case and the conflict is detected. 
This topic is also related to cases in which the \emph{methods holding the conflict are not adequately called}. 
In these cases, besides the proper creation of relevant objects, a sequence of method calls must be invoked with the aim of properly instantiating an object in such a way that the interference location can be reached.

\looseness=-1
\subsubsection*{Relevant Assertions exploring the Propagated Interference}
In a few cases, the generated tests even reach the interference location, infection occurs, but the test assertions do not properly explore the \emph{propagated} interference. 
For example, consider a conflict in the project CloudSlang,\footnote{This case refers to merge commit \href{https://github.com/CloudSlang/cloud-slang/commit/20bac30d9bd76569aa6a4fa1e8261c1a9b5e6f76}{20bac30} in project CloudSlang.} where the parent commits change the same \texttt{array}. 
The test case generated for this scenario correctly creates the object, calls the method in which the conflict occurs, and saves the method return object into a local variable. 
However, the generated assertion checks whether the local variable is \texttt{null} instead of exploring the object size. 
So, depending on the type of a method return object, the unit test generation tools could explore defined aspects that could detect the conflict. 
This way, tools could provide a list of handlers based on the types of objects under analysis.
For example, for array objects, these handlers would force the assertions to explore their size and contents.
For strings, assertions might explore comparisons between different strings, as also whether substrings are part of others, and so on.

\subsubsection*{Relevant Assertions relying on Interference Propagation} 
Test case assertions often explore the object returned by a method, but not objects that are passed as parameters. 
For example, again analyzing the changes performed of previous mentioned merge scenario of project Jsoup, the method \texttt{outerHtmlHead} requires three parameters as input, not returning any object (void method), as previously mentioned in this section. 
However, a semantic conflict occurs, and the first parameter holds the \emph{propagated} interference. 
The test case generated by EvoSuite focusses on verifying whether an exception is thrown during its execution. 
The assertions should not be restricted to explore objects returned by a method, but also other objects that are used by a method or any other way of communication.

%On a final note, the weaknesses of the tools we observed may be motivated by the diverse sample of real projects we adopt here. 
%Previous work assessing Randoop \,\cite{feedback-directed-randomPachecoShuvenduErnstBall2007}, for example, focuses mostly on generating tests for APIs. 
%For EvoSuite\,\cite{salahirad2019choosing}, previous work considers other kinds of projects, but many of them come from the same owner. 
%In other work \,\cite{fraser2014large}, the evaluation did not focus on detecting behavior changes.

\section{Threats to Validity}
\label{sec:validity}

%Our study design gives rise to some threats to validity that we mitigated as follows.
\emph{Construct Validity} 
As explained in Section~\ref{sec:motivating}, we cannot assess semantic conflict occurrence without having access to the developers' intentions or specification of the changes they make.
So our study focuses on interference occurrence.
As manually assessing global interference, and generating and running tests for the whole system, would demand considerable effort, our study is restricted to local interference occurrence. 
So the number of false negatives and false positives with respect to a global notion of interference could be different than what our results report. Nonetheless, regression tests could detect global interference if the interference is propagated, and if we generate tests for other classes in addition to the one that integrates the parallel changes made by two developers.

Aiming to increase the testability of the source code under analysis for the unit test generation tools, we apply testability transformations before performing our analysis. 
For example, we change access modifiers to \texttt{public}. 
Although this transformation breaks program encapsulation, it does not semantically change a program. 
If a semantic conflict can be observed accessing a class field, but this field is \texttt{private}, the unit test generation tools would face many problems trying to indirectly access this attribute without the transformation. 
Some may argue that, without this transformation, such conflict could never be observed. 
That might be true if indirect access, for instance with accessor methods, is not available, but we are aware of this and take it into consideration in our false positive analysis. 
Furthermore, the transformed program is only accessed by our semantic merge tool.

% update this part. The conflict was not detected because an abstract conflict holds the interference.
\emph{Internal Validity}
When creating the interference ground truth, we rely on manual analysis of unfamiliar source code.
%knowing the results of the test generation tools before the manual analysis could have influenced the verdict. 
To reduce this threat, we involve a group of six researchers, which are split in pairs during the analysis, and demand they provide an explanation of why there is no interference; this often requires understanding the changes in detail to detect refactorings, changed state elements, and how they impact each other.  
The risk is significantly reduced for the cases in which the tools are successful, as the threat can be minimized by analyzing the interference revealing test case, running it, and manually checking whether the test case assertions focus on the changed state elements.

\looseness=-1
%TODO2: tirei isso porque eh uma caracteristica da ferramenta, nao uma ameaca aos resultados. o que isso diz abaixo eh que uma outra versao de SAM poderia ter menos falsos negativos. ok, mas isso nao eh um threat. ou entendi errado? idem para a parte de flaky tests
% As we discuss when presenting our results, we remove from our analysis test cases that are not compilable on at least one commit of the merge scenario.  
% This limitation comes from the fact that test suites are generated for a specific parent commit, but has to be executed also in the \baseCommit END and \mergeCommit END commits. 
% However, if we could only consider the test case, there is a chance the conflict could be detected. 
% So the number of false negatives could be less than we report here if SAM handled that, but that does not . 
%We follow this proposal manually extending commits of a merge scenario and detecting the conflict, but with changes in the generated test case.

\looseness=-1
%Among the test cases we use to perform our study, there is a chance some of them are flaky.% tests, test cases that may either pass or fail without a specific reason. 
%We decide to remove these kinds of tests from our analysis as they could introduce both false positives and negatives in our results. 
%Nonetheless, these tests could also detect test conflicts if environmental or configuration problems cause the flakiness, and not the different merge scenario commits. 

\emph{External Validity} 
Our results are limited to the context of open-source GitHub Java projects. 
In the same way, the diversity of real projects we analyze here might have an impact on the detection of potential conflicts by our semantic tool.
The testability transformations, as we discuss, positively impact our results and contribute to increasing the source code testability; in some cases, also detecting the conflict. 
Applying our proposal of semantic merge tool to other programming languages would require test generation tools for the desired language and also the testability transformations, if applicable.

% !TeX spellcheck = en_GB
\section{Related work}\label{sec:related-work}
%\wm{Need to define: Semantic conflict, behavior change, regression testing, base, left.. illustrative figure}
%\tb{add references; add papers from slides}
\looseness=-1
\noindent
Regression testing has been used for detecting behavior changes in the past.
Evans and Savoia (\citeyear{evans2007differential}) combine regression and progressing testing (differential testing) to detect preserved, altered, and eliminated behavior of a program.
Jin et al. (\citeyear{jin2010automated}) present a test generator based on a list of changed classes between two versions of a program. 
Shamshiri et al. (\citeyear{shamshiri2013search}) present EvosuiteR, a test generation tool for differential testing that uses search-based algorithms to find regression faults on different versions of a program.
While the previous studies evaluate the detection of regression faults between two different versions of a program using regression tests, in this work, we evaluate the potential of regression tests to detect semantic conflicts on merge scenarios (three different versions of a program).
Campos et al. (\citeyear{campos2014continuous}) propose an approach (CTG) to more efficiently generate unit tests considering the whole project, instead of a single method or class as we do here. They focus on a \emph{Continuous Integration} context, and their approach can help detect behavior changes and regressions, as tests generated in a previous commit might yield a different result when executed in the next commit. But this is not enough for detecting interference as we do here, as our interference heuristic goes beyond behavior change detection; a test that was generated and passes in a parent commit, and that breaks in the following (say merge) commit, might indicate interference only if it breaks in the corresponding base commit. It would, however, be important to use CTG, instead of raw Evosuite focused on a target method as we do here, to assess whether it could improve SAM.
Hejderup and Gousios (\citeyear{hejderup2022can}) investigate the effectiveness of project test suites on detecting semantic changes motivated by updates to external dependencies instead of conflicting contributions applied to the project itself, as we do here. 
Although we do not investigate the occurrence of conflicts caused by conflicting changes involving external dependencies, the mentioned related work approach might be applied to detect these new conflicts.

\looseness=-1
Researchers also present techniques to detect and prevent conflicts early.
Palantir\,\citep{early-detection-conflicts-conservativeSarmaRedmilesHoek2012} is a workspace awareness tool that notifies developers of parallel changes in the same artifact. 
%An evaluation of the tool implied early detection and resolution of conflicts, fewer conflicts than before, and acceptable overhead. 
Brun et al. (\citeyear{early-detection-conflicts-speculativeBrunReidErnstNotkin2013}) propose incorporating speculative analysis for early detection and prevention of conflicts.
This way, they present Cristal, an assistive tool that compares remote and local individual collaborators' repositories in order to warn about possible code integration conflicts.
To detect test conflicts, they evaluate their technique by analyzing three Java projects and rely on project tests, which are often not enough for detecting interference as we explore here. 
The authors do not mitigate possible flaky tests in both studies, as we do in our study by executing the test suites multiple times.
The failed tests are not executed on the parent and base commits of the merge scenario, as we do here, which may result in false positives, as the failed test may occur due to the changes exclusively performed by one parent.
These studies have also investigated ways in which conflicts can be prevented early, thereby minimizing their impact on productivity. 

Cavalcanti et al. (\citeyear{improved-semistructured-mergeCavalcantiBorbaAccioly2017,semistructured-merge-jsTavaresBorbaCavalcantiSoares2019,semistructured-vs-structured-mergeCavalcantiBorbaSeibtApel2019}) conduct empirical studies that analyze merge scenarios and compare the accuracy of different merge resolution techniques: unstructured, semi-structured, and structured merge. 
They also propose a new semi-structured tool with significant advantages over unstructured merge tools by reducing the false-positive and false-negative rates of earlier semi-structured tools. 
%\tb{did they do the comparison and is it a finding of their work, or is this insight from somewhere else?}
Overall, they find that exploring more structure does not necessarily improve merge accuracy. 
Contrasting with our investigation here, 
%\tb{our work? and contrasting what exactly? their empirical study, their proposed technique, or their methodology?},
their proposed tools 
%\tb{the tools they evaluate?}
are not able to detect behavioral semantic conflicts, only syntactic and static semantics conflicts.
%\tb{this sentence is not clear to me}.

Nguyen et al. (\citeyear{nguyen2015detecting}) present Semex, a tool for detecting which combination of merged changes 
%\tb{unify terminology how you refer to the merged code; currently using different terms, which is confusing}
causes a test conflict based on a technique called variability-aware execution \,\citep{nguyen2014exploring}. %\tb{is Semex a tool for integrating variants in the first place? how is it beneficial for our integrations of two variants?}
First, the tool separates the changes done by each parent commit in the merge scenario and encodes each one using conditionals around them (\texttt{if} statements) to integrate all these changes in a single program.
%\tb{what does conditionals to encode mean?}
Semex then uses variability-aware execution to detect semantic conflicts by running existing project tests, if available, on this single program, 
%\tb{on the single program?}
exploring all possible combinations of the encoded changes. %\tb{where do the project tests come from? from the different variants? how can Semex using the tests explore all possible combinations?}
Reporting a conflict exclusively based on the failure of a test in the merged code does not always imply a conflict or interference. If the test fails in one of the parent commits too, 
%\tb{what's that?}
failure in the merge might simply indicate inheritance of a defect. 
%\tb{very hard to follow}
%\textcolor{blue}{(This paper is explained in greater depth than the others. 2-line summary should be enough. Also, the next lines where Semex is differentiated with our approach can be written more concisely)}
That is why we propose different criteria, based on the idea of tests as partial specifications of the changes to be integrated.  
We also rely on and assess the use of test generation tools to detect conflicts, instead of relying on existing project tests, which are often missing or have limitations, as described above. 

Due to restricted sample sizes, related work on semantic merge tools does not discuss precision and recall measures.
\rev{Wuensche et al. (\citeyear{wuensche2020detecting}) suggest an approach based on static analysis and a tool to detect and predict the occurrence of test conflicts, as they formally call \emph{higher-order merge conflicts}. 
Based on the changes performed during a merge scenario, they (re)build a call graph and detect potential dependencies among merge scenario code fragments that lead to a conflict.
To detect test conflicts, the authors manually analyze build records of merge scenarios and extract change patterns that lead to test conflicts based on the authors' observation.
As a result, 22 potential conflicts out of 1489 merge scenarios are reported by the tool. 
To validate the potential conflicts, the authors search for bugs reported after each merge scenario occurrence.
However, they do not confirm conflict occurrences as no bug is reported.}

\looseness=-1
Sousa et al. (\citeyear{semantic-merge-verificationSousaDillingLahiri2018}) propose \textit{SafeMerge}, a tool that leverages compositional verification to check \textit{semantic conflict freedom} in merge scenarios. 
In principle, this kind of static analysis should lead to more false positives and fewer false negatives, when compared to the use of tests as we propose here. 
An evaluation with 52 merge scenarios indicates that \textit{SafeMerge} reports  75\% of the scenarios without conflicts, with a false positive rate of 15\%.
However, analyzing the merge scenarios reported with conflicts, we conclude that some of them do not represent conflicts according to our criteria. 
In these cases, the changes involved do not interfere with each other or are only refactorings, leading to no behavior change and consequently no interference.
Due to the experimental design and dataset characteristics, the authors do not present false negative rates. However, concerning false positives, the authors disclose a rate of 3.8\% (2 out of 52 cases), whereas our study demonstrates a rate of 3.5\% (3 out of 85 cases).

Arcuri and Galeotti (\citeyear{arcuri2021enhancing}) adopt a related approach by presenting a set of testability transformations; unlike our transformations, they do not focus on changing code element access modifiers or semantic changes but support and guide the search algorithm when generating tests. 
Their core idea is based on \emph{Method Replacements}, which replace specific method calls at the bytecode level with their customized methods. 
To evaluate their technique, they implement it as an extension of the EvoMaster tool and perform an empirical study analyzing ten Rest web service projects (open-source and industrial ones). 
The results show that the techniques effectively improve code coverage and fault detection. 

Tiwari et al. (\citeyear{tiwari2021production}) present PANKITI, a related approach regarding the use of serialization to support unit test tools. 
Unlike our approach to serializing objects based on a target method during merge scenarios, PANKITI monitors an application in production serializing objects when target methods are called. 
While we serialize the current objects holding the target method and its required parameters, the authors also serialize the returning target method objects. 
For our context, we are not interested in returning objects as we focus on objects that might let us reach infection states of conflicting contributions. 
Furthermore, infections are not always propagated through returning target method objects; in our sample, we observe infections being propagated through parameters, for example. 

Regarding the issues we observe by the tools when generating test suites, previous studies also bring evidence about the hardness of dealing with complex objects~\citep{fraser20151600,silva2017analyzing,da-silva2020detecting}.
These related studies discuss the difficulty of generating complex objects, which are required when calling specific methods under analysis.
By complex objects, we consider objects with multiple other objects from internal as also external dependencies.
In order to address these issues, we propose feeding the tools with serialized objects leading them to reuse previous objects originally created by the original project test suite.

\section{Conclusion}\label{sec:conclusion}
\looseness=-1
\noindent

\looseness=-1
In this work, we present and evaluate a semantic merge conflict detection technique using automated test-case generation. 
As opposed to prior attempts in the literature, our strategy does not require explicitly defined behavior specifications or substantial setup effort.
We define interference criteria and systematically investigate their effectiveness by detecting conflicts upon a manually curated ground-truth dataset originating from 85 changes' pairs from 51 software merge scenarios that integrate changes to the same method, constructor, or field declaration mined from GitHub.

\looseness=-1
In order to detect conflicts, we combine unit test generation tools and adopt improvements, like testability and serialization transformations. %\tb{the term is not clear here; explain that these tools were specifically improved for this purpose, and that we investigated their impact (did we?)} in the source code to be analyzed.
As a result, we show the feasibility of a semantic merge tool, SAM (SemAntic Merge tool). 
While SAM is able to detect nine conflicts out of 28 conflicts from 85 changes' pairs, we report only three false positives according to our interference criteria. 
This suggests that semantic merge tools based on unit test generation would help developers detect semantic conflicts early, otherwise reaching end-users as failures.
Regarding the unit test tools, our results show that SAM best performs when combining only the tests generated by Differential EvoSuite and EvoSuite.
The Testability Transformations improve the testability of target code under analysis in three of the nine detected interference cases, suggesting that they might be useful for interference detection. 
We discuss necessary improvements to test generation and make our manually curated dataset available in a replication package~\citep{appendix-paper2}, also to help building future semantic merge tool.

\looseness=-1
We also explore and measure the impact of different improvements in our semantic merge-conflict detection technique due to the limitations of unit test tools and the complexity of the target code under analysis.
First, we propose and evaluate the use of serialized objects as input for the tools during the generation process.
Although we do not observe new semantic conflicts detected after applying this technique, new general behavior changes are detected involving pairs of commits.
Second, we also extend Randoop aiming to maximize the number of tests exploring the target method, if applicable.
Although our tool Randoop Clean reports better results regarding test suites dealing with more diverse objects, we do not observe the detection of new conflicts.
As future work, we plan to extend SAM to consider original project test suites to detect conflicts based on our conflict criteria, improve the use of serialization transformations and extend unit test tools.

\section*{CRediT authorship contribution statement}
\noindent
\textbf{Leuson Da Silva}: Conceptualization, Methodology, Investigation, Data curation, Writing – original draft. \textbf{Paulo Borba}: Conceptualization, Methodology, Supervision, Writing – original draft. \textbf{Toni Maciel}: Investigation, Data curation. \textbf{Wardah Mahmood}: Investigation, Writing – review \& editing. \textbf{Thorsten Berger}: Supervision, Writing – review \& editing. \textbf{João Moisakis, Aldiberge Gomes, Vinicius Leite}: Investigation.

\section*{Declaration of competing interest}
 The authors declare that they have no known competing financial interests or personal relationships that could have appeared to influence the work reported in this paper.

\section*{Acknowledgements}
\label{sec:acknowledgement}
\noindent
We thank Marcelo d'Amorim, Rohit Gheyi, Leonardo Fernandes, Breno Miranda, Leonardo Murta, and the anonymous reviewers for valuable comments to improve an earlier version of this paper.
We thank Rafael Alves, Galileu Santos, Matheus Barbosa, and Thaís Burity for their support when creating our dataset.
We also thank INES (National Software Engineering Institute), the Brazilian research funding agencies CNPq (309741/2013-0), FACEPE (IBPG-0692-1.03/17 and APQ/0388-1.03/14), and CAPES, as well as the Swedish Research Council (257822902), Vinnova Sweden (2016-02804) and the Wallenberg Academy.







%\section*{References}

\bibliography{elsarticle-template-harv}

%\begin{thebibliography}{00}



%\end{thebibliography}
\end{document}